\documentclass[pdflatex,sn-mathphys-num]{sn-jnl}

\usepackage{graphicx}
\usepackage{amsmath,amssymb,amsfonts}%
\usepackage{amsthm}%
\usepackage{mathrsfs}%
\usepackage[title]{appendix}%
\usepackage{bm}
\usepackage{acronym}
\usepackage{xspace}
\usepackage{xcolor}
\usepackage[shortlabels]{enumitem}
\usepackage{hyperref}
\usepackage[capitalize]{cleveref}
\usepackage{caption}
\usepackage{subcaption}
\usepackage{multirow}
\usepackage{array}
\usepackage[symbol]{footmisc}
\usepackage{booktabs}
\usepackage[normalem]{ulem}
\usepackage{aas_macros} 

\acrodef{GW}{gravitational wave}
\acrodef{BH}{black hole}
\acrodef{BBH}{binary black hole}
\acrodef{PE}{parameter estimation}
\acrodef{LVK}{LIGO-Virgo-KAGRA}
\acrodef{RJ MCMC}{reversible-jump Markov chain Monte Carlo}
\acrodef{PDF}{probability density function}
\acrodef{CDF}{cumulative density function}
\acrodef{PPD}{posterior predictive distribution}
\acrodef{GWTC-4}{4th Gravitational Wave Transient Catalog}
\acrodef{ZAMS}{zero age main sequence}
\acrodef{SMT}{stable mass transfer}
\acrodef{CHE}{chemically homogeneous evolution}
\acrodef{IMF}{initial mass function}
\acrodef{FAR}{False Alarm Rate}

\acrodef{PISN}{pair-instability supernova}
\acrodefplural{PISN}[PISNe]{pair-instability supernovae}

\acrodef{AGN}{active galactic nucleus}
\acrodefplural{AGN}[AGN]{active galactic nuclei}

\newcommand\di{\ensuremath{\mathrm{d}}\xspace}
\newcommand\deriv[2]{\ensuremath{\frac{\mathrm{d}#1}{\mathrm{d}#2}}\xspace}

\newcommand\vectheta{\ensuremath{\bm{\theta}}\xspace}

\newcommand\vecLam{\ensuremath{\mathbf{\Lambda}}\xspace}
\newcommand\veclam{\ensuremath{\bm{\lambda}}\xspace}
\newcommand\veclams{\ensuremath{\{\bm{\lambda}\}}\xspace}

\newcommand\PL{\ensuremath{\mathrm{PL}}\xspace}
\newcommand\chieff{\ensuremath{\chi_{\mathrm{eff}}}\xspace}
\newcommand\mmin{\ensuremath{m_{\mathrm{min}}}\xspace}
\newcommand\mmax{\ensuremath{m_{\mathrm{max}}}\xspace}
\newcommand\mtmax{\ensuremath{m_{2, \mathrm{max}}}\xspace}
\newcommand\Msun{\ensuremath{\mathrm{M}_\odot}\xspace}

\newcommand\BF{\ensuremath{\mathcal{B}}\xspace}

\DeclareUnicodeCharacter{03C7}{\ensuremath{\chi}}

\newcommand{\appropto}{\mathrel{\vcenter{
  \offinterlineskip\halign{\hfil$##$\cr
    \propto\cr\noalign{\kern2pt}\sim\cr\noalign{\kern-2pt}}}}}

\newcommand\chieffhightailskewt{\ensuremath{0.46^{+0.17}_{-0.12}}\xspace}

\newcommand\chiefflowtailskewt{\ensuremath{0.00^{+0.29}_{-0.40}}\xspace}

\newcommand\chieffmedtailskewt{\ensuremath{0.23^{+0.17}_{-0.18}}\xspace}
\newcommand\chieffmedtenskewt{\ensuremath{0.05^{+0.04}_{-0.04}}\xspace}
\newcommand\chieffmedthirtyskewt{\ensuremath{0.00^{+0.04}_{-0.04}}\xspace}

\newcommand\chiefflowtailNPLNP{\ensuremath{-0.18^{+0.27}_{-0.30}}\xspace}

\newcommand\chieffmedtailNPLNP{\ensuremath{0.11^{+0.16}_{-0.12}}\xspace}
\newcommand\chieffmedtenNPLNP{\ensuremath{0.04^{+0.03}_{-0.03}}\xspace}
\newcommand\chieffmedthirtyNPLNP{\ensuremath{-0.01^{+0.03}_{-0.04}}\xspace}

\newcommand\muchiposconfidencetenNPLNP{\ensuremath{98.6\%}\xspace}
\newcommand\muchiposconfidencetenskewt{\ensuremath{98.2\%}\xspace}

\newcommand\mumtenskewt{\ensuremath{10.3^{+1.4}_{-1.4}}\xspace}
\newcommand\mumthirtyskewt{\ensuremath{23.7^{+10.2}_{-13.3}}\xspace}

\newcommand\mumtenNPLNP{\ensuremath{10.0^{+1.0}_{-0.9}}\xspace}
\newcommand\mumthirtyNPLNP{\ensuremath{28.1^{+3.6}_{-2.8}}\xspace}

\newcommand\muqtenskewt{\ensuremath{0.55^{+0.38}_{-0.40}}\xspace}

\newcommand\muqthirtyskewttenperc{\ensuremath{0.71}\xspace}

\newcommand\muqthirtyNPLNPtenperc{\ensuremath{0.76}\xspace}

\newcommand\bfractailskewt{\ensuremath{0.02^{+0.08}_{-0.02}}\xspace}
\newcommand\bfractenskewt{\ensuremath{0.77^{+0.12}_{-0.12}}\xspace}
\newcommand\bfracthirtyskewt{\ensuremath{0.19^{+0.12}_{-0.11}}\xspace}

\newcommand\bfractailNPLNP{\ensuremath{0.11^{+0.20}_{-0.08}}\xspace}
\newcommand\bfractenNPLNP{\ensuremath{0.77^{+0.08}_{-0.18}}\xspace}
\newcommand\bfracthirtyNPLNP{\ensuremath{0.12^{+0.07}_{-0.05}}\xspace}

\newcommand\ratetailskewt{\ensuremath{0.5^{+1.6}_{-0.5}}\xspace}
\newcommand\ratetenskewt{\ensuremath{14.6^{+7.0}_{-4.9}}\xspace}
\newcommand\ratethirtyskewt{\ensuremath{3.7^{+2.6}_{-2.1}}\xspace}

\newcommand\ratetailNPLNP{\ensuremath{2.1^{+4.3}_{-1.5}}\xspace}
\newcommand\ratetenNPLNP{\ensuremath{14.7^{+6.7}_{-5.5}}\xspace}
\newcommand\ratethirtyNPLNP{\ensuremath{2.3^{+1.5}_{-1.0}}\xspace}

\newcommand\gammaskewt{\ensuremath{3.0^{+0.9}_{-1.2}}\xspace}

\newcommand\gammatentwocomp{\ensuremath{5.7^{+2.1}_{-2.8}}\xspace}

\newcommand\gammatenbiggerconfidenceAB{\ensuremath{94.6\%}\xspace}
\newcommand\gammatenbiggerconfidenceABC{\ensuremath{92.2\%}\xspace}

\newcommand\rhothirty{\ensuremath{0.80^{+0.13}_{-0.44}}\xspace}

\newcommand\qmaxskewtIDten{\ensuremath{0.80^{+0.13}_{-0.16}}\xspace}

\DeclareCaptionType{extdata}
\DeclareCaptionLabelFormat{extdata}{\textbf{Extended Data Fig.~#2}}
\crefname{extdata}{Extended Data Fig.}{Extended Data Figs.}
\Crefname{extdata}{Extended Data Fig.}{Extended Data Figs.}

\makeatletter
\newenvironment{extdatafigure}
  {\begin{figure}%
   \def\@captype{extdata}%
   \captionsetup{labelformat=extdata,labelsep=space}}
  {\end{figure}}
\makeatother

\begin{document}

\title{Reversible-jump MCMC reveals binary black hole subpopulations with distinct redshift evolution}


\author*[1]{\fnm{April Qiu} \sur{Cheng}}
\email{aqcheng@princeton.edu}

\author[2,3]{\fnm{Alexandre} \sur{Toubiana}}
\email{alexandre.toubiana@unimib.it}

\author[4]{\fnm{Sylvia} \sur{Biscoveanu}}
\email{sbisco@princeton.edu}

\author[5]{\fnm{Jonathan} \sur{Gair}}
\email{jonathan.gair@aei.mpg.de}

\affil*[1]{\orgdiv{Department of Astrophysical Sciences}, \orgname{Princeton University}, \orgaddress{\street{4 Ivy Lane}, \city{Princeton}, \state{NJ} \postcode{08544}, \country{USA}}}

\affil[2]{\orgdiv{Dipartimento di Fisica “G. Occhialini”}, \orgname{Universitá degli Studi di Milano-Bicocca}, \orgaddress{\street{Piazza della Scienza 3}, \city{Milano}, \postcode{20126}, \country{Italy}}}

\affil[3]{\orgname{INFN, Sezione di Milano-Bicocca}, \orgaddress{\street{Universitá degli Studi di Milano-Bicocca}, \city{Milano}, \postcode{20126},  \country{Italy}}}

\affil[4]{\orgdiv{Department of Physics}, \orgname{Princeton University}, \orgaddress{\city{Princeton}, \state{NJ} \postcode{08544}, \country{USA}}}

\affil[5]{\orgdiv{Astrophysical and Cosmological Relativity}, \orgname{Max Planck Institute for Gravitational Physics (Albert Einstein Institute)}, \orgaddress{\street{Am Mühlenberg 1}, \city{Potsdam}, \postcode{14476}, \country{Germany}}}

\date{\today}

\abstract{Analyses of the growing catalog of \acf{BBH} mergers observed by the LIGO-Virgo-KAGRA detectors are beginning to resolve features in their population-level mass, spin, and redshift distributions, revealing imprints of the astrophysical processes driving their formation and evolution. We present a novel method to search for subpopulations in the data using reversible-jump Markov chain Monte Carlo, providing interpretable results while making minimal prior assumptions. We find evidence for three subpopulations: a narrow subpopulation in primary mass at $\sim 10\,\Msun$ with preferentially aligned spins and unequal masses, consistent with isolated binary evolution; a subpopulation broadly distributed around $\sim 30\,\Msun$ with isotropically-distributed spins and a strong preference for equal mass ratios, consistent with dynamical formation in clusters; and a high-spin subpopulation spanning the continuum in mass, which we interpret as the confluence of multiple subdominant formation channels. When we allow for the independent redshift evolution of each subpopulation, we find that the subpopulation encompassing the $10\,\Msun$ peak evolves more quickly than the $30\,\Msun$ subpopulation, with implications for the delay-time distribution and metallicity-dependent \ac{BBH} formation efficiency. Our work lays the foundation for a novel data-driven framework to infer the formation mechanisms of \acp{BBH}.
}

\maketitle

\section{Introduction}

Among the most salient open questions in astrophysics is the formation mechanism(s) of \acfp{BBH}. Different astrophysical formation channels are expected to produce \acp{BBH} with distinct signatures in the multidimensional parameter space defined by their masses, spin vectors, and redshifts. For example, tides in the stellar progenitors of mergers of isolated binaries are expected to efficiently align the progenitor spins to the orbital angular momentum, leading to component black holes with preferentially aligned spin vectors~\cite{1993MNRAS.260..675T, 2000ApJ...541..319K, 2004PhRvD..69j2002G, 2014LRR....17....3P, 2016Natur.534..512B, 2016MNRAS.458.2634M, 2016A&A...588A..50M, 2016ApJ...832L...2R, 2017NatCo...814906S}; gas accretion will also align (or anti-align) the spin vectors in mergers of \acp{BH} trapped in the disks of \acp{AGN}~\cite{2018ApJ...866...66M, 2019PhRvL.123r1101Y, 2020ApJ...899...26T, 2020MNRAS.494.1203M, 2024A&A...685A..51V}. On the other hand, mergers of \acp{BH} from dynamical interactions in a dense cluster are expected to have isotropically distributed spin orientations~\cite{1993Natur.364..423S, 2009ApJ...692..917M, 2010ARA&A..48..431P, 2013LRR....16....4B, 2016ApJ...832L...2R}, while those formed in stellar triples may have spins preferentially perpendicular to the orbital plane due to a combination of precession and Kozai--Lidov effects driven by the outer body~\cite{2018ApJ...863...68L, 2018MNRAS.480L..58A, 2018ApJ...863....7R, 2020PhRvD.102l3009Y}. For reviews of possible \ac{BBH} formation pathways and their observational signatures, see e.g. \cite{2022LRR....25....1M,2025arXiv250203523B,2022PhR...955....1M,2021hgwa.bookE..16M}.

We are now in a position to address this question using \ac{GW} observations of coalescing \acp{BBH} by the \acf{LVK} Collaboration \cite{2015CQGra..32g4001L,2025PhRvD.111f2002C,2025CQGra..42h5016S,2015CQGra..32b4001A,2021PTEP.2021eA101A}. The release of the \acf{GWTC-4} \cite{Collaboration2025} has more than doubled the catalog size compared to its predecessor, providing significant insight into the structure of the underlying source population \cite{2025arXiv250818083T}. There is now increasing evidence for multiple subpopulations of \acp{BBH}, each with distinct mass, mass ratio, and spin properties, consistent with a population sourced by a mixture of formation channels \cite{2025arXiv250915646B,2026arXiv260317987R,2025arXiv251122093S}.

Methods for looking for subpopulations in the data can be roughly grouped into \textit{strongly-modeled} and \textit{weakly-modeled} approaches. In the former, one typically describes the population as a mixture 
of astrophysically-motivated parametric functional forms, each representing a hypothesized formation channel \cite{2026arXiv260107908P,2021ApJ...910..152Z,2026arXiv260317987R,2021ApJ...915L..35K,2022ApJ...941L..39W,2024PhRvL.133e1401L,2024A&A...692A..80P,2025arXiv250915646B,2025arXiv251122093S,2025arXiv251105316T,2025arXiv250923897L}, and infers their relative rates; one can also fit the data directly to the output of population synthesis models \cite{2021ApJ...910..152Z,2025ApJ...988..189C,2023ApJ...950..181W}. While such strongly parameterized models offer direct astrophysical interpretability, they risk imposing structure that is absent from the data, or hiding structure that is not anticipated by the model \cite{2023ApJ...955..127C}. On the other hand, weakly-modeled approaches make only minimal assumptions about the population distribution (e.g., that it is smooth), instead letting the data drive the shape of the inferred population \cite{2024PhRvX..14b1005C,2023ApJ...946...16E,2023ApJ...957...37R,2025arXiv251122093S,2023ApJ...957...37R,2021CQGra..38o5007T,2025PhRvD.111f3043H,2022MNRAS.509.5454R,2023MNRAS.524.5844T,2023PhRvR...5b3013P,2024PhRvL.133e1401L,2025arXiv250909123A,2025arXiv250620731A,2026arXiv260414290G}. The increased flexibility comes at the cost of additional model and computational complexity and reduced astrophysical interpretability. In particular, the identification of subpopulations typically requires some heuristic extra processing steps, such as dividing the population in parameter space by eye \cite{2025arXiv251122093S,2026arXiv260414290G}.

In this work, we present a novel ``best of both worlds'' search for subpopulations using \acf{RJ MCMC} \cite{1995Biome..82..711G}, which simultaneously performs Bayesian model comparison between models of varying complexity along with inference on the model parameters. 
\ac{RJ MCMC} has previously been applied to the analysis of the \ac{BBH} population by \cite{2023MNRAS.524.5844T}, which used reversible-jump to optimize the complexity of one-dimensional semi-parametric and non-parametric models, and later by \cite{2025ApJ...994L..52T}, which extended it to multidimensional non-parametric inference.
This work is motivated by the following idea: because the number of subpopulations within the \ac{BBH} population is unknown, \ac{RJ MCMC} is a natural and computationally efficient framework to identify them without over-fitting the data.
That is, for a mixture model of $n$ suitably flexible subpopulations, \ac{RJ MCMC} finds the best $n$ along with the features of each subpopulation. 
Because our model still consists of a mixture of physical model components, rather than a purely descriptive representation of the data, we retain the advantage of interpretability over non-parametric methods.
In short, this is a data-driven method with the clarity of a parametric approach.

Applying this framework to the \ac{GWTC-4} data, we robustly identify three subpopulations consistent across multiple variants of population models. We find 1) a subpopulation at $m_1 \approx 10\,\Msun$ with small positive \chieff\ and a mild preference for unequal-mass pairings, consistent with isolated binary evolution; 2) a broad peak at $m_1 \sim 30\,\Msun$ with a $\chieff$ distribution centered at zero and a strong preference for equal-mass pairings, consistent with dynamical assembly; and 3) a high-spin continuum extending to high masses with a positive-mean, broad $\chieff$ distribution, which includes the exceptional events of the catalog. 
We also find evidence for differences in the evolution of different subpopulations with redshift, implying a short delay-time distribution and inefficient binary progenitor formation above moderate metallicities for the subpopulation including the $10~\Msun$ peak and the continuum.

\section{Three binary black hole subpopulations}

Population inference of \ac{GW} data is typically performed using hierarchical Bayesian inference, in which one fits a population model to the source parameters $\vectheta$ representing an individual binary in the catalog, which are in turn inferred from the \ac{GW} strain data. 
The key astrophysical ingredient is the population model, which we briefly describe below; details of our methodology can be found in \cref{sec:methods}.


We model the \ac{BBH} population as a mixture of $n$ components on the 4-dimensional parameter space $\vectheta=[m_1, q, \chieff, z]$ consisting of the primary mass, mass ratio, effective aligned spin, and redshift, such that the differential merger rate per unit detector time $t$ is given by 
\begin{equation}  \label{eq:general_diff_rate}
\begin{split}
\frac{\di N}{\di \vectheta \di t}(\veclams, \vecLam) &= \sum_{k=1}^{n} \frac{\di N_k}{\di t \di \vectheta}(\veclam_k, \vecLam) \\
    &= \deriv{V_c}{z} \sum_{k=1}^{n} \frac{ \mathcal{R}_{0, k} \psi(z | \vecLam)}{1+z}   p_k(m_1, q, \chieff | \veclam_k, \vecLam),
\end{split}
\end{equation}
where each subpopulation $k$ has a local ($z=0$) merger rate $\mathcal{R}_{0, k}$ and a distribution $p_k(\vectheta | \veclam_k, \vecLam)$ dependent on its subpopulation parameters $\veclam_k$ as well as global population parameters $\vecLam$ shared between all subpopulations. For example, in our default population model we model the source-frame merger rate as a power-law
\begin{equation}
    \psi(z|\gamma) = (1+z)^\gamma
\end{equation}
with index $\gamma \in \vecLam$ shared between all subpopulations. This assumption can be relaxed, the results of which we will discuss in \cref{sec:correlations}.
In our framework, $n$ is a free parameter; \ac{RJ MCMC} returns posterior samples on $\veclams$ and $\vecLam$ for all models of different $n$, with the number of samples proportional to the relative evidences, or equivalently, the Bayes factors (\BF). 

To verify the robustness of our results, we search for subpopulations across several variants of population models, which differ primarily in the functional form of the mass distribution.
In our default population model (\texttt{skewt}), we model the primary mass distribution of each subpopulation as a skew-t distribution, which is a flexible, 4-parameter distribution that can accommodate both asymptotic power-law and Gaussian behavior, and can broadly capture the features of most smooth, unimodal distributions. We compare these results to a mixture model of power-laws and Gaussians (\texttt{NPLNP}), which is similar to common strongly parameterized models in the literature \cite{2018ApJ...856..173T,2025arXiv250818083T}, except here the number of components is not fixed. 
In both of these cases, we model the conditional mass ratio distribution $p(q|m_1)$ as a truncated Gaussian.
We also employ a population model with an alternate parameterization of the mass ratio (\texttt{skewt\_ID}), in which both \acp{BH} in the binary are drawn from the same \ac{BH} mass distribution. Their pairing is described by a correlation coefficient $-1 < \rho < 1$, where $\rho=0$ describes independent \ac{BH} pairing, and $\rho=1$ and $\rho=-1$ are extremes where the masses are perfectly correlated (i.e., mass ratio is always equal) and perfectly anti-correlated (in quantile, i.e., the \acp{CDF} evaluated at the two masses always sum to $1$), respectively. 
This parameterization has the advantage of reduced complexity compared to the $2$ parameters of a Gaussian. 

\begin{figure}
\centering
\includegraphics[width=\textwidth]{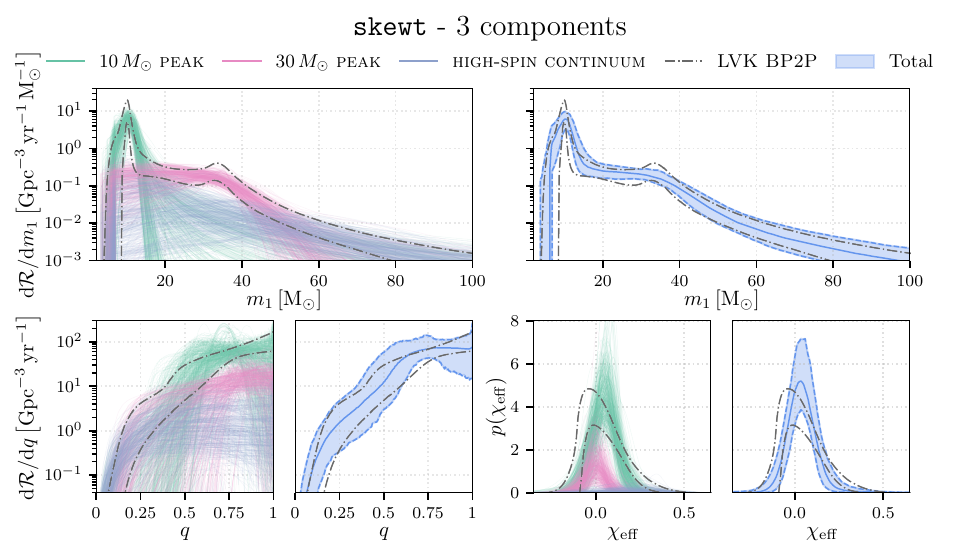}
\caption{\label{fig:skewt_PPDs} Posterior sample draws for differential merger rates as a function of primary mass, mass ratio and effective spin at $z=0.2$ in our fiducial \texttt{skewt} 3-component model. The $10\,\Msun$ peak, $30\,\Msun$ peak, and high-spin continuum subpopulations are plotted in green, pink, and indigo, respectively, while the posterior median and $90\%$ credible intervals of their total are plotted in shaded blue on the right for each parameter. The corresponding curves for the \textsc{Broken power-law + 2 Peaks} population model of the \acs{LVK} analysis \cite{2025arXiv250818083T}, which uses a skew-normal model for \chieff and a power-law distribution for the mass ratio, are plotted for comparison (dashed-dot gray).}
\end{figure}


\begin{figure}
\centering
\includegraphics[width=\textwidth]{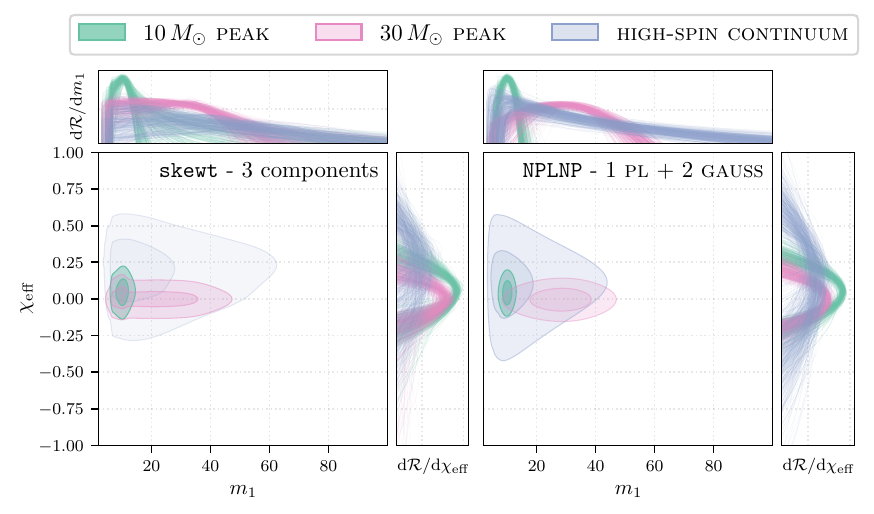}
\caption{\label{fig:m1_chieff_contours} $50\%$ and $90\%$ contours of the median posterior $\di \mathcal{R}^k / \di m_1 \di \chieff$ for each subpopulation $k$, evaluated at $z=0.2$; opacity of the shading corresponds to the relative rates of the subpopulations. We show the most preferred model of the \texttt{skewt} (left) and \texttt{NPLNP} (right) population models, with 3 subpopulations in both. Corresponding posterior sample draws from the marginalized 1D distributions (in log-scale) are shown in the top and side panels.}
\end{figure}

\begin{figure}
\centering
\includegraphics[width=\textwidth]{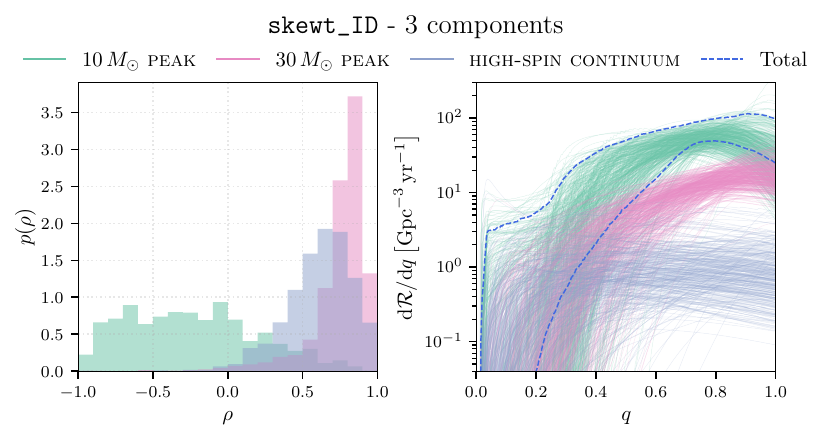}
\caption{\label{fig:rho_hist} Posteriors of the mass pairing correlation coefficient $\rho$ for each subpopulation (left) and the corresponding differential merger rate as a function of $q$ at $z=0.2$, inferred from the \texttt{skewt\_ID} population model. In the right panel, colored curves are random posterior draws of the marginal differential merger rate on $q$ for each subpopulation, with the $90\%$ credible interval of the total distribution plotted in dashed blue. The $10\,\Msun$ peak is consistent with uncorrelated pairing, resulting in a mass ratio distribution peaked away from $q=1$, while the $30\,\Msun$ peak shows a strong preference for equal mass ratios.}
\end{figure}

For each population model, we show results from the most preferred model, which is 3 components for \texttt{skewt} and \texttt{skewt\_ID}, and 2 power-laws + 1 Gaussian for \texttt{NPLNP}. As we discuss further below, this model is only weakly preferred over a 2-component model. Bayes factors between the models of different complexity within each population model can be found in \cref{tab:bfs}.

\cref{fig:skewt_PPDs} shows draws from the 1D marginalized distributions of each subpopulation as well as their total, inferred from our default \texttt{skewt} population model; the corresponding results for the \texttt{skewt\_ID} and \texttt{NPLNP} population models can be found in \cref{fig:all_ppds_other_models}.
The marginalized distributions are largely consistent with the \ac{LVK} analysis \cite{2025arXiv250818083T}. In $m_1$, we recover a sharp peak at $10\,\Msun$, a bump around $30\,\Msun$, and a power-law-like tail at high masses. While our \texttt{NPLNP} analysis prefers 1 power-law + 2 Gaussians compared to the preferred \textsc{Broken Power-Law + 2 Peaks} model in the \ac{LVK} analysis, this difference is attributable to the fact that including an extra power-law component at the break incurs a heavier Occam's razor penalty, as each subpopulation has independent mass ratio and spin distributions~\cite{2025arXiv250818083T}. Our inferred mass ratio distribution is consistent with peaking at $q\sim0.7-1$, while the \ac{LVK} results more strongly favor a mass ratio distribution that peaks at equal masses. In contrast to the \ac{LVK} \chieff population model, which specifically allows and finds evidence for skewness~\cite{2025ApJ...990..147B}, our population model uses a symmetric \chieff distribution, so such an asymmetry would need to be constructed from the overlap of multiple subpopulations.
Where subpopulations overlap, there is greater degeneracy between the relative contributions from each subpopulation, as shown by the fact that the total population distribution is better constrained than the distributions of each subpopulation.

The $3$ identified subpopulations are consistent across all population models that we test. 
We compare the results of the \texttt{skewt} and \texttt{NPLNP} population models in \cref{fig:m1_chieff_contours}. Here, we plot the posterior differential merger rate as a function of $m_1$ and \chieff, the two best measured parameters in the parameter space $\vectheta$; different subpopulations occupy visibly distinct regions of parameter space.
The subtle differences between the population models illustrate the impact of strongly parameterized models. Comparing the two $m_1$ distributions in the top insets, we can see that the \texttt{NPLNP} model forces the high-spin continuum subpopulation to peak at lower masses due to its parameterization as a power-law, and forces the $30\,\Msun$ peak to drop off at small masses due to its parameterization as a Gaussian. In contrast, inference with the \texttt{skewt} population model finds significant support around $m_1 \lesssim 10$ for the $30\,\Msun$ subpopulation. In turn, this causes subtle differences in the resulting \chieff and mass ratio distributions of each subpopulation.
Nonetheless, the fact that both population models are still qualitatively consistent lends credibility to our results. 

We now describe the three subpopulations as follows:

\bmhead{$10\,\Msun$ peak}
This is a narrow subpopulation centered at $m_1 \approx 10\,\Msun$ with small positive \chieff, consistent with isolated formation channels producing \acp{BBH} with spins aligned with the orbital angular momentum. We plot this subpopulation in green throughout this paper. 

The peak of the primary mass distribution is\footnote{In this paper, all quoted uncertainties correspond to the median and $90\%$ central credible intervals unless otherwise specified.} $\mumtenskewt\,\Msun$ ($\mumtenNPLNP\,\Msun$) in the \texttt{skewt} (\texttt{NPLNP}) population model. 
The mean of the \chieff distribution is $\mu_\chi=\chieffmedtenskewt$ (\chieffmedtenNPLNP), with $\mu_\chi>0$ measured at \muchiposconfidencetenskewt (\muchiposconfidencetenNPLNP) credibility. 
This is also the largest subpopulation of the three, with a branching fraction of \bfractenskewt (\bfractenNPLNP) and a merger rate of \ratetenskewt (\ratetenNPLNP) $\rm{Gpc}^{-3}\,\rm{yr}^{-1}$.

Because the masses of \acp{BBH} in this subpopulation are confined within a narrow mass range, mass ratios are not well measured. 
Nonetheless, there is tentative evidence that this subpopulation supports unequal mass ratios more than the rest of the population. 
As seen in the left panel of \cref{fig:rho_hist}, this subpopulation shows the strongest support for independent pairing of \ac{BH} masses, with $\rho < 0.38$ with 90\% credibility in the \texttt{skewt\_ID} population model.
The induced marginal $p(q)$ distribution, shown in the right panel, peaks at $q_{\rm peak}= \qmaxskewtIDten < 1$.
However, this signature is weaker in the default population model, possibly due to the increased model complexity: the posterior on the peak of the conditional mass ratio distribution $\mu_q$ deviates only weakly from the prior, although still consistent with unequal mass ratios ($\mu_q=\muqtenskewt$). 

The properties of this subpopulation---a narrow mass peak, a small but robustly positive mean \chieff, a preference for unequal pairings, and the bulk of the overall merger rate---are collectively consistent with an isolated field binary evolution origin. 
The small positive mean \chieff\ we infer is consistent with a general isolated binary formation scenario: spins are generally expected to be aligned with the orbital angular momentum, and tidal spin-up of the tidally-locked second-born progenitor naturally produces small but positive $\chieff$, even if the first-born \ac{BH} is born slowly spinning \cite{2020A&A...635A..97B,2018A&A...616A..28Q,2023ApJ...952...53M,2019ApJ...881L...1F}. 

A peak at $10\,\Msun$ has been found previously in the literature \cite{2021ApJ...913L..19T,2023ApJ...946...16E,2023PhRvX..13a1048A,2026arXiv260401420L} and interpreted in several isolated formation scenarios. 
The \ac{SMT} formation scenario predicts a peak around this mass range, as it is inefficient at producing mergers below $\sim9-10\,\Msun$. This is because orbits must shrink via the direct loss of mass carrying away angular momentum during the second mass-transfer phase, as opposed to drag forces present in a binary common envelope during dynamically unstable mass transfer, and such orbital tightening is inefficient for lower mass \acp{BBH} \cite{2022ApJ...940..184V,2023ApJ...948..105V}. \ac{SMT} can also explain the preference for unequal masses that we find for this subpopulation, as unequal mass \acp{BBH} are more likely to shrink their orbits during the mass transfer episode~\cite{2024A&A...689A.305O}, with mass ratios predicted to peak around $q \sim 0.6-0.8$ \cite{2019MNRAS.490.3740N,2022ApJ...931...17V}. A peak around $9\,\Msun$ has also been predicted in the mass spectrum of \acp{BH} formed from binary-stripped stars. Core compactness reaches a local maximum in that mass range as a consequence of a transition to neutrino-dominated burning, causing decreased explodability and hence increased probability of \ac{BH} formation \cite{2025A&A...694A.186G,2023ApJ...950L...9S,2025arXiv251007573W}. Finally, ref. \cite{2026arXiv260401420L} targed the $10\,\Msun$ peak as a separate subpopulation, interpreting the peak as failed-supernova \acp{BH} forming from direct collapse in a narrow progenitor-mass range \cite{2024ApJ...964L..16B}. They also find a mass ratio and \chieff distribution of the subpopulation different from the rest of the population, consistent with our findings.

\bmhead{$30\,\Msun$ peak}

This subpopulation has $m_1$ broadly distributed around $30\,\Msun$, a \chieff distribution centered at $0$, and a strong preference for equal mass ratios, consistent with dynamical formation scenarios in clusters.  We plot this subpopulation in pink throughout this paper.

Quantitatively, this subpopulation peaks in primary mass at \mumthirtyskewt (\mumthirtyNPLNP) in the \texttt{skewt} (\texttt{NPLNP}) population model, although in \texttt{skewt} the peak is more akin to a plateau with a fall-off starting around $\sim 30\,\Msun$, as seen in the top left panel of \cref{fig:skewt_PPDs}. 
The mean of the \chieff distribution is squarely zero, with $\mu_\chi=\chieffmedthirtyskewt$ (\chieffmedthirtyNPLNP). \cref{fig:rho_hist} shows that this subpopulation has the strongest preference for equal mass pairings out of all the subpopulations, with $\rho=\rhothirty\sim1$ in the \texttt{skewt\_ID} population model and $\mu_q>\muqthirtyskewttenperc$ (\muqthirtyNPLNPtenperc) in \texttt{skewt} (\texttt{NPLNP}) at 90\% credibility. This is the second-most populous subpopulation, with a branching fraction of \bfracthirtyskewt (\bfracthirtyNPLNP) and merger rate of \ratethirtyskewt (\ratethirtyNPLNP)~$\rm{Gpc}^{-3}\,\rm{yr}^{-1}$.

Isotropic spin orientations are a robust signature of formation scenarios where mergers are formed dynamically in a globular or nuclear star cluster, which implies a $\chieff$ distribution symmetric about zero \cite{2021NatAs...5..749G}. 
Furthermore, mass segregation and the larger interaction cross-section of heavier \acp{BH} both favor dynamical pairings at high mass and near-equal mass ratios \cite{2016ApJ...824L...8R,2017MNRAS.469.4665P}.
Moreover, some simulations of dense star clusters are able to predict the overdensity around $30\,\Msun$ as well as the measured merger rate \cite{2023MNRAS.522..466A,2025A&A...701A.252B,2026arXiv260320430A}, given sufficiently high cluster densities. 
While first-generation mergers in \ac{AGN} disks are also predicted to have a \chieff distribution centered at zero \cite{2025ApJ...993..163C}, they do not predict a strong preference for equal mass ratios.

The pile-up of \acp{BBH} around this mass-range has been well-known even from early phenomenological analyses of the \ac{BBH} mass spectrum \cite{2018ApJ...856..173T, 2019ApJ...882L..24A}.
Other analyses have also found the equal-mass and/or isotropic spin properties of the \acp{BBH} in this mass range, including both flexible  \cite{2025arXiv251122093S} and strongly-parameterized mixture models \cite{2025arXiv250915646B,2026arXiv260317987R} and interpreted this subpopulation as dynamically assembled.

\bmhead{High-spin continuum}

This is a small subpopulation spanning the continuum in primary mass, including a tail out to high masses, which we plot in indigo throughout this paper. Note that the existence of this subpopulation is only mildly preferred by the data when it is modeled by a skew-t distribution compared to a 2-component model, with $\BF=1.3$ ($2.7$) in the \texttt{skewt} (\texttt{skewt\_ID}) population models; see \cref{tab:bfs}. In the 2-component model, this continuum is absorbed into the peak at $10\,\Msun$. However, an analogous 2-component model is disfavored in the \texttt{NPLNP} analysis with $\BF=0.08$, which could be a consequence of the decreased flexibility compared to the \texttt{skewt} population model.

Due to the wide range of masses spanned by this subpopulation, it naturally allows for the most extreme mass ratios. As seen in \cref{fig:m1_chieff_contours}, it also has a much broader \chieff distribution than the other two subpopulations centered at a significantly positive value. 
Quantitatively, the \chieff distribution is centered at $\mu_\chi=\chieffmedtailskewt$ (\chieffmedtailNPLNP) in the \texttt{skewt} (\texttt{NPLNP}) population model. It is also useful to examine the 90\% central interval spanned by the \chieff Gaussian; we find $\chieff^{5\%}=\chiefflowtailskewt$ (\chiefflowtailNPLNP) and $\chieff^{95\%}=\chieffhightailskewt$. Thus, while a \chieff distribution with entirely non-negative support is neither significantly favored nor ruled out, this subpopulation confidently supports very positive values of $\chieff \sim 0.4$.
This is the smallest subpopulation, with a branching fraction of \bfractailskewt (\bfractailNPLNP) and a local merger rate of \ratetailskewt (\ratetailNPLNP) $\rm{Gpc}^{-3}\,\rm{yr}^{-1}$. 

Unlike the other two subpopulations, this subpopulation resists an obvious astrophysical interpretation. 
Hierarchical mergers in stellar clusters, where one or both of the \acp{BH} are formed from a previous merger, do produce higher-mass \acp{BBH} and large spins, but the expectation is for $\chieff$ to be symmetric about zero with $|\chieff| \lesssim 0.5$ \cite{2016ApJ...832L...2R,2019PhRvD.100d3027R,2025PhRvL.134a1401A,2021NatAs...5..749G} due to the prediction for the \ac{BH} spin orientations to be isotropic with respect to the orbital angular momentum.
Instead, we find that $\mu_\chi>0$ with $99\%$ ($93\%$) credibility.
This is in contrast with the zero-centered distributions inferred for the high-mass \acp{BBH} in \cite{2025arXiv250904637A,2026arXiv260107908P,2026arXiv260317987R}, who have interpreted those \acp{BBH} as hierarchically merged in clusters.
As we will discuss in \cref{ssec:events}, while many events identified as hierarchical merger candidates in previous works do fall into this subpopulation, the \chieff distribution of this subpopulation is inconsistent with consisting entirely of cluster hierarchical mergers.
Several other formation channels therefore plausibly contribute to this subpopulation. The inferred properties of this subpopulation are consistent with the \ac{CHE} of low-metallicity stars, predicted to produce massive, near-equal-mass \acp{BBH} with high positive aligned spins \cite{2016A&A...588A..50M,2016MNRAS.458.2634M,2016MNRAS.460.3545D,2025ApJ...995L..76P}, as well as hierarchical mergers in the disks of \ac{AGN}, where gas torques can partially align spins and produce a $\chieff$ distribution skewed to positive values  \cite{2017ApJ...835..165B,2020ApJ...899...26T,2025arXiv250923897L,2025ApJ...989...67D}.
A high-mass subpopulation with aligned spins was found in \cite{2025arXiv250923897L} and also interpreted as hierarchical mergers in \ac{AGN} disks.
We therefore interpret this subpopulation as a residual ``catch-all'' component that naturally absorbs diverse exceptional events in our framework, rather than as the signature of any single formation channel. 
Further observations, and possibly the inclusion of more spin parameters in our analysis, may be able to resolve the formation channels that constitute this subpopulation.

Besides its spin properties, another indication of a hierarchical origin for a \ac{BBH} is if it populates the \ac{BH} mass gap expected from \acp{PISN}. Recently, ref. \cite{2025arXiv250904151T} has found evidence for the \ac{PISN} mass gap in the secondary mass distribution. 
Using a variant of our default population model, we investigate and verify the presence of a sharp drop-off in the secondary mass distribution. The results can be found in \cref{fig:m2max}.



\section{Individual events} \label{ssec:events}

\begin{figure}
\centering
\includegraphics[width=0.75\textwidth]{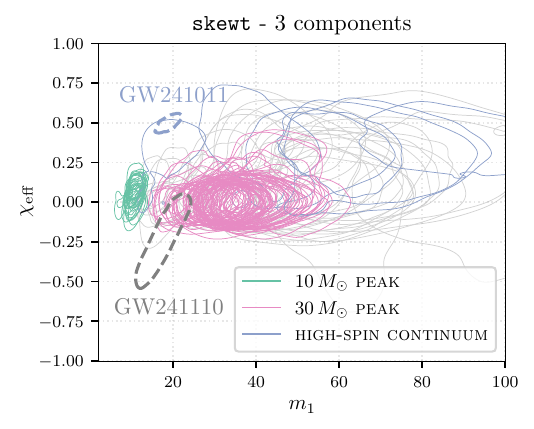}
\caption{\label{fig:skewt_pwp} 90\% contours in primary mass and effective spin of the population-weighted posteriors of the $153$ \ac{GWTC-4} events along with GW241011 and GW241110, using the results of our default \texttt{skewt} analysis. Events with $>75\%$ probability of belonging to a subpopulation are color-coded accordingly, and otherwise plotted in gray. The clustering of the three subpopulations is clearly visible.}
\end{figure}

Next, we examine the posterior probability that each individual observed event belongs in each subpopulation, and where each event is located in parameter space.
\cref{fig:skewt_pwp} shows the default \texttt{skewt} 3-component model's population-weighted posteriors of all analyzed events in $m_1$ and \chieff. Events which are likely ($>75\%$) to belong to one subpopulation are color-coded accordingly; otherwise, they are plotted in gray.

\begin{figure}[p]
\centering
\includegraphics[width=\textwidth]{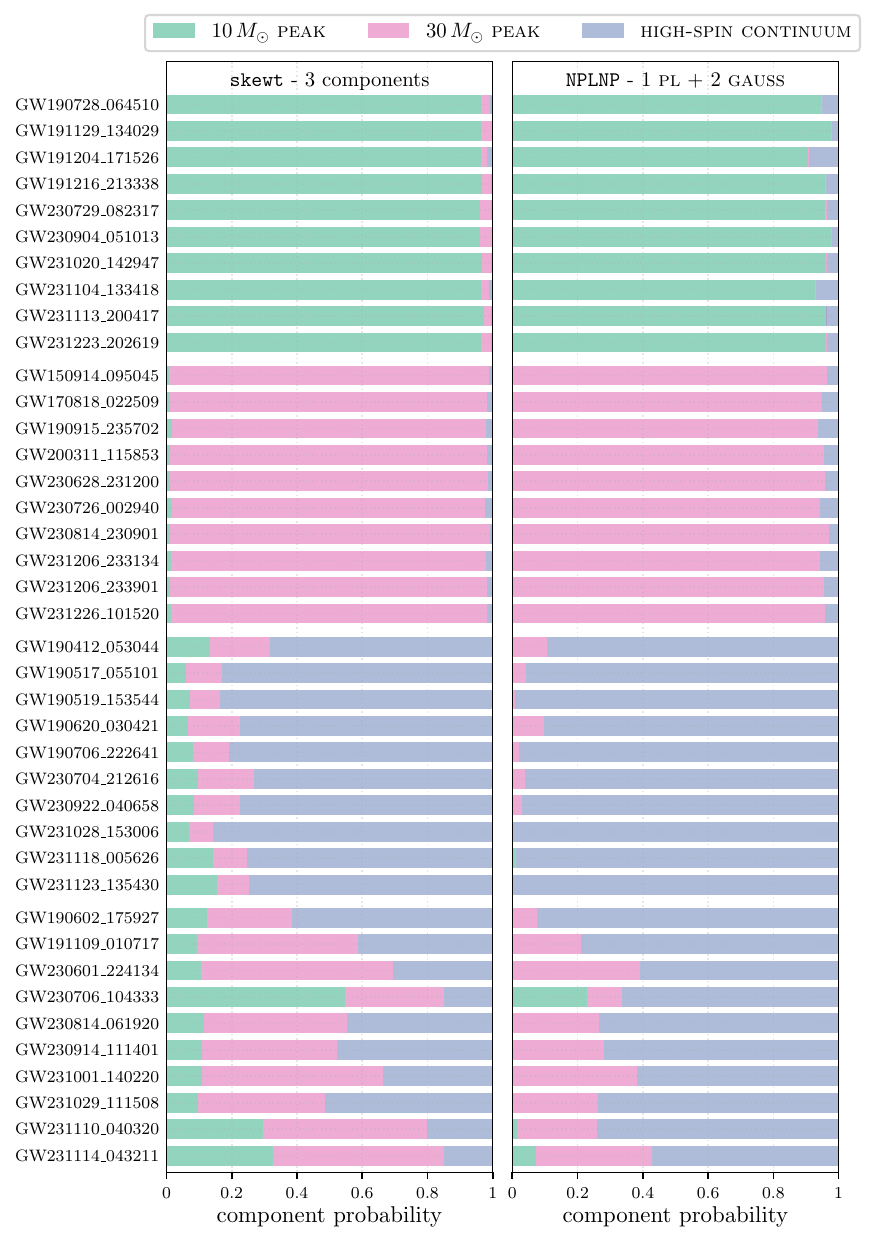}
\caption{\label{fig:subpop_prob} Probabilities of individual events belonging to each subpopulation for the \texttt{skewt} 3-component model (left) and \texttt{NPLNP} 1 power-law + 1 Gaussian model (right). We plot in blocks of $10$ the top events of each \texttt{skewt} subpopulation, as well as the model's most ambiguous events at the bottom, which we define as the events with the smallest Kullback-Leibler divergence from a uniform distribution between the 3 subpopulations. Events are organized chronologically within each block.}
\end{figure}

We identify these events in \cref{fig:subpop_prob}, which shows the default model's subpopulation probabilities of the top 10 most confident events in each subpopulation, as well as the top 10 most ambiguous events.
We then compare this to the subpopulation probabilities of the \texttt{NPLNP} 1 power-law + 2 Gaussians model of the same events, which we plot to the right. 
All events that are confidently detected in each subpopulation in the \texttt{skewt} population model are also confidently detected in the \texttt{NPLNP} model, supporting the robustness of these subpopulations. 

The high-spin continuum subpopulation contains many of the exceptional events in the catalog in at least one of our population models.
These include GW190412~\cite{2020PhRvD.102d3015A}, which is notable for its highly asymmetric masses, as well as GW231123~\cite{2025ApJ...993L..25A}, which is both notable for their high masses and spins.
GW190412 has also been identified as a candidate for an \ac{AGN} disk-driven hierarchical merger origin \cite{2021ApJ...908..194T,2022MNRAS.517.5827F,2020PhRvL.124y1102G}, and both \ac{CHE} \cite{2025ApJ...995L..76P} and hierarchical \ac{AGN}-disk mergers \cite{2026ApJ...999..127L,2025arXiv250813412D} have been proposed as formation scenarios for GW231123.
Additionally, the primary spin of GW190412 is too low \cite{2020ApJ...900..177K} and the spins of GW231123 too high \cite{2025ApJ...992L..26S} for a cluster hierarchical merger scenario.
Nonetheless the high-spin continuum subpopulation does contain most of the events identified as candidates for hierarchical mergers in previous works, including GW190517\_055101, GW231028\_153006 \cite{2026arXiv260107908P}, GW231118\_005626 \cite{2025arXiv251105316T}, GW190519\_153544, GW190620\_030421, and GW190706\_222641 \cite{2021ApJ...915L..35K}. 
This further emphasizes the ambiguous astrophysical interpretation of this subpopulation.

We can also examine the population-weighted posteriors of the O4b exceptional events GW241011 and GW241110~\cite{2025ApJ...993L..21A,Collaboration2025a}, which we show in \cref{fig:skewt_pwp}. These events are identified as the most likely candidates for containing a \ac{BH} with a hierarchical origin \cite{2025ApJ...993L..21A,2026arXiv260107908P,2025arXiv251105316T}, although we do not include them in our population analysis (as doing so would constitute essentially adding two events to the population by hand). 
With the \texttt{skewt} (\texttt{NPLNP}) population model, we find that the probability that GW241011 belongs in the high-spin continuum is $90.4\%$ ($99.9996\%$), while the corresponding probability for GW241110 is $35\%$ ($91\%$).
This is consistent with the ambiguous astrophysical interpretation of this subpopulation, especially in the \texttt{skewt} population model.
It is important to note that because both events have exceptional spins, including them in the population analysis has the potential to significantly change the properties of this subpopulation.
Furthermore, because we only use \chieff rather than the full component spin vectors, we are not sensitive to more specific predictions of the hierarchical merger channel, such as the expectation for one or more of the component \acp{BH} to have spin magnitude $\chi \approx 0.7$ \cite{2005PhRvL..95l1101P,2008PhRvD..77b6004B}.
Reapplying our framework to the full O4b catalog, and possibly expanding to more spin parameters, will constitute an important next step for finding a robust subpopulation of hierarchical mergers.

\section{Correlations} \label{sec:correlations}

\subsection{Different subpopulations evolve differently in redshift}

\begin{figure}
\centering
\includegraphics[width=\textwidth]{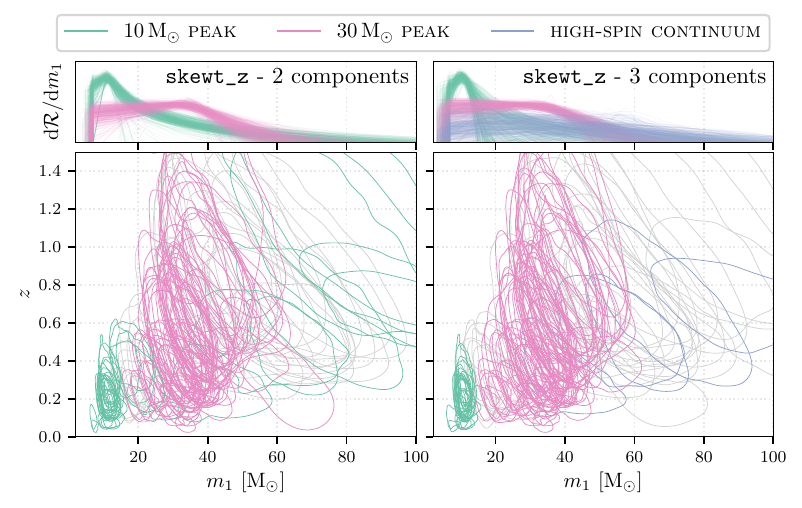}
\caption{\label{fig:skewt_z_pwp} Population-weighted posteriors color-coded by subpopulation as in \cref{fig:skewt_pwp}, except here we plot $m_1$ and $z$ of the 2-component (left) and 3-component (right) models of the \texttt{skewt\_z} population model. The top inset shows posterior draws of the rate of each subpopulation at $z=0.2$ as a function of $m_1$. Because the continuum is included in the $10\,\Msun$ peak in the 2-component model, the subpopulation includes events at high masses and high redshifts.}
\end{figure}

In our default model, we assumed a global rate evolution $\psi(z|\gamma)$, where $\gamma$ is the global power-law index, shared between all subpopulations. We relax this assumption in the \texttt{skewt\_z} population model, allowing each subpopulation to have its own $\gamma$.
We recover the same subpopulations as above, except now the 2-component model is slightly favored over the 3-component model with $\BF = 1.6$, likely due to the increase in the parameter space. The top panels of \cref{fig:skewt_z_pwp} show draws from the marginalized $m_1$ population distributions of both models. As before, the $10\,\Msun$ peak combines with the high-spin continuum in the 2-component model, whereas the 3-component model distributions are statistically consistent with those of the default \texttt{skewt} population model. 

\begin{figure}
\centering
\includegraphics[width=4.5in]{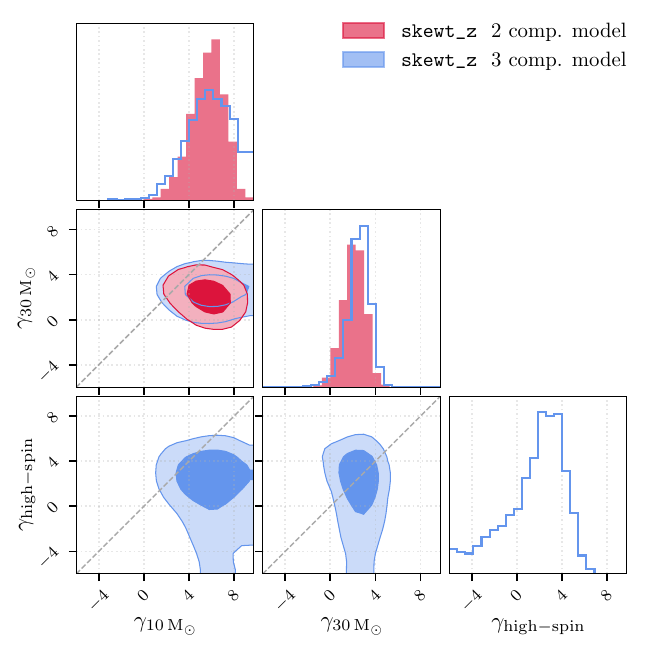}
\caption{\label{fig:gamma_hist} Posteriors on the power-law index $\gamma$ of the source-frame rate evolution with redshift of our different subpopulations for the 2-component model (red) and the 3-component \texttt{skewt\_z} model. The diagonal gray line demarcates where $\gamma_A=\gamma_B$ for any subpopulations $A, B$. In both models, the $10\,\Msun$ peak subpopulation prefers a faster rate evolution than the other component(s).}
\end{figure}

\cref{fig:gamma_hist} shows a corner plot on $\gamma$ of each subpopulation for both the 2-component and 3-component models. We find that the $10\,\Msun$ peak favors a significantly faster rate evolution, with $\gamma=5.7^{+2.1}_{-2.8}$ ($6.0^{+3.2}_{-3.7}$) in the 2-component (3-component) model, compared to $\gamma = \gammaskewt$ of the \texttt{skewt} inference which assumes a global rate evolution, consistent with the \ac{LVK}'s inference. Compared to the $30\,\Msun$ peak subpopulation, $\gamma_{10\,\Msun} > \gamma_{30\,\Msun}$ at \gammatenbiggerconfidenceAB (\gammatenbiggerconfidenceABC) credibility.


The main implication of this result is that the $10\,\Msun$ peak and continuum will dominate as we go to higher redshifts with respect to the $30\,\Msun$ peak.
Of course, this claim is only robust if events belonging to the $10\,\Msun$ peak provide information at high redshifts.
It is therefore instructive to inspect the population-weighted posteriors of the events, which we show in \cref{fig:skewt_z_pwp}. 
As seen in the left panel, the $10\,\Msun$ peak subpopulation is still informed by events at higher redshifts due to the fact that it extends to the high-mass continuum in the 2-component model, and therefore events around $10\,\Msun$ as well as events in the continuum have evidence for a fast redshift evolution.
Indeed, evidence for a faster redshift evolution of a subpopulation of high-mass, highly-spinning \acp{BBH} has been found in previous works using more strongly parameterized models tailored to search for specific correlations~\cite{2024ApJ...975...54G,2026ApJ..1001L..40F}. On the other hand, the faster redshift evolution of a re-analysis of selected low mass events have also found a faster redshift evolution~\cite{2026arXiv260414290G}.

\begin{figure}
\centering
\includegraphics[width=0.9\textwidth]{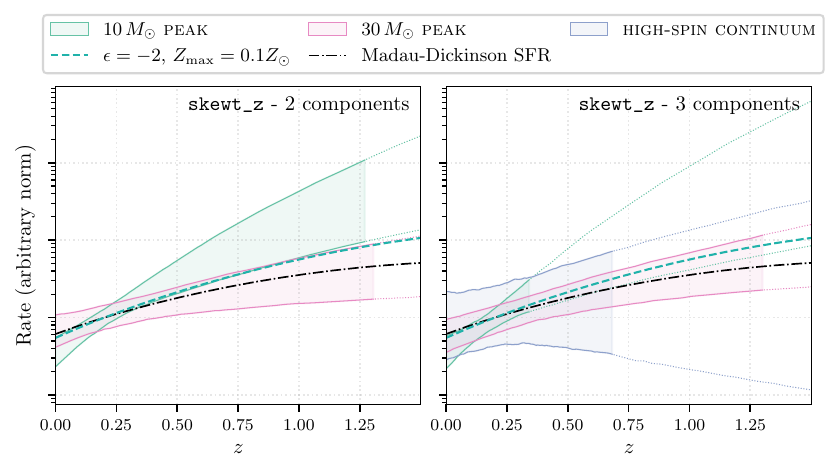}
\caption{\label{fig:sfr} Source-frame \ac{BBH} merger rate as a function of redshift for each subpopulation, as well as the theoretical prediction for our simplified model with $p(t_d)\propto t_d^\epsilon$ and inefficient \ac{BBH} formation above $Z_{\max}$ (turquoise dashed line). The dash-dot black line shows our fiducial Madau-Dickinson star formation history. Posterior draws from the population model are normalized to their median rate at $z=0.2$, and the theoretical merger history is normalized to its value at $z=0.2$. As described in the text, the truncation of the shaded regions corresponds to the redshift beyond which the subpopulation has no redshift information from the data.}
\end{figure}

Expanding on this idea, we can define a ``maximum informative redshift" of each subpopulation as the greatest median population-weighted posterior redshift of any event which likely ($>75\%$) belongs in the subpopulation. In \cref{fig:sfr}, we plot the source-frame redshift evolution of each subpopulation, truncating the shading at the maximum informative redshift of each subpopulation.
We show the resulting $m_1$ and \chieff distributions at different redshifts in \cref{fig:ppd_evolution}.

The fast redshift evolution of the $10\,\Msun$ peak is nontrivial to explain with the standard picture of the \ac{BBH} merger rate following the star formation history with some delay time. Following \cite{2024ApJ...967..142T}, we consider a model in which the \ac{BBH} merger rate follows the Madau-Dickinson star formation history $\psi_{\rm MD}(z)$ \cite{2014ARA&A..52..415M} with a power-law delay-time distribution $p(t_d) \propto t_d^\epsilon$ and inefficient \ac{BBH} formation above a cutoff metallicity $Z_{\max}$. Thus, the \ac{BBH} source-frame birth rate of \ac{BH} binaries is proportional to $\psi_{\rm MD}(z) \, F(Z_{\max}; z)$, where $F(Z_{\max}; z)$ is the fraction of star formation occurring below metallicity $Z_{\max}$, adopting the analytic prescription in ref. \cite{2006ApJ...638L..63L}.

We test a grid of combinations of $\epsilon$ and $Z_{\max}$ and compare the predicted rate evolution to the rate evolution inferred for our subpopulations.
Because $\gamma\approx6$ is significantly faster than the best-fit star-formation history (which has $\gamma=2.7$ and a turnover at redshift $1.9$, well beyond the majority of events in the catalog), we find that attempting to match the constraints inferred from the preferred two-component model requires adopting extremely short delay-time distributions together with a maximum metallicity for efficient \ac{BBH} formation well below solar
($Z_{\max} \lesssim 0.1~Z_\odot$). Even in these relatively extreme scenarios, however, we still struggle to reproduce the inferred constraints for $z \gtrsim 1$.
This is illustrated by the dashed red line in \cref{fig:sfr}, showing the rate computed from $\epsilon=2$ and $Z_{\max}=0.1~Z_\odot$, which in the 2-component model (left panel) is barely consistent with the lower edge of the 90\% credible interval of the $10\,\Msun$ peak + continuum subpopulation. Slower delay times, introducing some minimum $t_d$, or a larger $Z_{\max}$ all predict a rate evolution inconsistent with that of the $10\,\Msun$ peak and continuum.

In the 3-component model (right panel), although a steeper evolution for the $10\,\Msun$ peak is still preferred, the posteriors on $\gamma$ are broader, due to the increase in model complexity and a decrease in the number of events per subpopulation. Thus, the power-law index for the  $10\,\Msun$ peak is compatible with that of the  $30\,\Msun$ peak within error bars, and in this model no constraints are placed on the delay time distribution or cutoff metallicity, especially since the faster rate evolution of the $10\,\Msun$ peak is confined to low redshifts of $z\lesssim0.3$. 

\subsection{Correlations including spins}

Previous works have found a broadening of the \chieff distribution with redshift \cite{2022ApJ...932L..19B,2026ApJ..1001L..40F} and a correlation between mass ratio and \chieff \cite{2022MNRAS.517.3928A,2023ApJ...958...13A,2026ApJ...999L..30V}, although evidence for the latter has become more ambiguous in the latest dataset \cite{2025arXiv250818083T}. We find no strong evidence for either of these correlations in our analysis, which we show in \cref{fig:q_chieff_contours,fig:ppd_evolution}.
Our results suggest that the broadening of the \chieff distribution with redshift is either driven by an intrinsic evolution of the \ac{BH} spin distribution, or by subpopulations with more complex redshift evolution than can be accommodated by a power law (e.g. subpopulations localized within a narrower range of redshifts). 
This stands in contrast with the results of ref. \cite{2026ApJ..1001L..40F}; we note, however, that their analysis included the two remarkable O4b events GW241110 and GW241011, and employed a population model specifically designed to target a subpopulation of hierarchical mergers.


\section{Conclusion}

The origin of \acp{BBH} is one of the central questions in modern astrophysics. Active theoretical efforts are ongoing to answer this question, and a wide range of formation channels have been proposed. These include isolated binary evolution involving orbital tightening through stable or unstable mass transfer, dynamical assembly in dense stellar clusters or triple systems, \ac{CHE} in close, low-metallicity binaries, hierarchical mergers of previous merger remnants, gas-assisted mergers in the disks of \acp{AGN}, and more. The joint distributions of masses, spins, and redshift predicted by these models differ, and therefore provide a means to disentangle the relative contributions of each formation channel.
With $\mathcal{O}(10^2)$ confidently detected mergers in \ac{GWTC-4}, features in this multidimensional parameter space are beginning to be resolved with enough granularity to begin addressing this question directly from the data, by asking which and how many subpopulations the data prefers.

Using a novel method to search for subpopulations within the \ac{BBH} population leveraging \ac{RJ MCMC}, we find evidence for three subpopulations with distinct mass, spin, and redshift:
(i) a narrow concentration at $m_1 \approx 10\,\Msun$ with small positive \chieff, a preference for unequal pairings, and the bulk of the merger rate, consistent with isolated field binary evolution including the \ac{SMT} channel; (ii) a broad peak at $m_1 \sim 30\,\Msun$ with a $\chieff$ distribution centered on zero and a strong preference for equal-mass pairings, consistent with dynamical assembly in dense stellar clusters; and (iii) a continuum extending from small to large masses with a positive-mean, broad $\chieff$ distribution, which contains the exceptional high-mass and high-spin events of the catalog but is inconsistent with originating exclusively from cluster hierarchical mergers, for which the $\chieff$ distribution should be symmetric about zero. The positive mean we recover hints at contributions from \ac{CHE} or \ac{AGN}-disk hierarchical mergers, and is plausibly explained by contributions from multiple channels.

We also find the first data-driven evidence for a faster evolution of the $10\,\Msun$ peak with redshift with respect to the $30\,\Msun$ peak, with the power-law rate index $\gamma_{10\,\Msun} > \gamma_{30\,\Msun}$ at $\gammatenbiggerconfidenceAB$ credibility in the most preferred model. Furthermore, the inferred value $\gamma_{10\,\Msun}=\gammatentwocomp$ implies short delay-time distributions and inefficient \ac{BBH} formation at metallicities above $Z_{\max} \lesssim 0.1\,Z_\odot$, assuming a Madau-Dickinson star formation history. Understanding the astrophysical implications of this result more rigorously will be an interesting next step for future work.


Previous works have identified subpopulations with distinct primary mass, mass ratio, and spin properties \cite{2025arXiv250915646B,2026arXiv260317987R}.
For example, the $3$-subpopulation mixture model of ref. \cite{2026arXiv260317987R} finds roughly the same three subpopulations as our analysis, with similar astrophysical interpretations.
One significant difference, however, is that the \chieff distribution of their broad continuum subpopulation is centered at $0$ simply due to the fact that it was a targeted search for a hierarchical subpopulation, and thus the symmetry of the \chieff distribution about $0$ was enforced in the model prior. As this example illustrates, our framework has the advantage of being more agnostic than strongly-modeled analyses.
Thus, we are able to verify their results and draw equally robust astrophysical conclusions with far more minimal prior assumptions, with the added benefit of not needing to manually compare the evidence for models of varying complexity.
We also propose that moving forward, more flexible parametric distributions, such as the skew-t distribution, should be used to resolve features in the \ac{BBH} population.

In conclusion, we demonstrate that it is possible to resolve robust, astrophysically-interpretable features in the \ac{BBH} population without strong prior astrophysical assumptions imprinted in the model. 
Two of the three identified subpopulations map cleanly to an astrophysical formation channel, and the asymmetric spin signature of the high-spin continuum component is itself an informative result, and one that strongly parameterized analyses tailored to a hierarchical interpretation are not free to recover. 
Furthermore, our results can already confidently attribute most of the events in the catalog to a subpopulation, and therefore a possible formation channel. 
Upcoming catalogs will have source counts approaching $\mathcal{O}(10^3)$~\cite{2018LRR....21....3A} as well as better-measured source parameters, including the component spins. 
We expect that the inclusion of spin parameters beyond \chieff, along with a larger catalog, will be able to resolve the high-spin continuum component into its constituent formation channels. It should also sharpen the constraints on the delay-time distribution and the metallicity dependence of \ac{BBH} formation, and allow us to identify the formation channel of individual events with much higher confidence. The long-standing question of the origins of \acp{BBH} is, at last, becoming an empirical one.


\section{Methodology} \label{sec:methods}

In this section, we explain the technical details of our methodology. We describe the basics of hierarchical Bayesian inference of \ac{GW} data in \cref{ssec:hbi}, and the details of our population models in \cref{ssec:popmodels}, including a population model in which we verify the existence of a secondary mass gap as found in ref. \cite{2025arXiv250904151T}. Then, we describe \ac{RJ MCMC} in \cref{sec:rjmcmc} and the implementation details in \cref{ssec:implementation}. 
In \cref{ssec:popweights}, we describe how we computed the population-weighted posteriors and subpopulation probabilities reported in \cref{ssec:events}. Finally, in \cref{ssec:unique} we describe our approach to sorting the identified subpopulations. 

\subsection{Hierarchical Bayesian inference} \label{ssec:hbi}


Given a catalog of observed \acp{BBH}, we are interested in inferring the population properties from which these \acp{BBH} arise. One can do this by means of hierarchical Bayesian inference. Recalling \cref{eq:general_diff_rate}, given some population model $\di N / \di\vectheta /\di t$ governed by subpopulation parameters $\veclam$ and global population parameters $\vecLam$, one can infer the posterior distribution on $\veclam, \vecLam$ given the data. This inference is hierarchical because the population is informed by the observed distribution of source parameters $\vectheta=[m_1, q, \chieff, z]$, which are in turn inferred from the \ac{GW} strain data for each event via a Bayesian inference step called \ac{PE}. Therefore, one must marginalize over $\vectheta$ for each event. The likelihood of $N$ observed \acp{BBH} $\{d\}$ under selection effects is given by an inhomogeneous Poisson likelihood \cite{2019MNRAS.486.1086M,2022hgwa.bookE..45V,2020PASA...37...36T}
\begin{equation} \label{eq:hbi_ll}
    \mathcal{L}\left(\left\{d\right\}|\veclams,\vecLam \right) \propto  e^{-\xi(\veclams,\vecLam) N(\veclams,\vecLam)} \prod_{i=1}^{N_{\text {obs }}} \int \frac{\mathrm{d} N}{\mathrm{~d} \vectheta}(\veclams,\vecLam) \frac{p_{\text{PE}}(\vectheta | d_i)}{\pi_{\text{PE}}(\vectheta)} \mathrm{d} \vectheta
\end{equation}
where $p_{\text{PE}}\left(\vectheta | d_i\right)$ and $\pi_{\text{PE}}(\vectheta)$ are the \ac{PE} posterior and prior, respectively, and thus we can evaluate \cref{eq:hbi_ll} via Monte Carlo integration over the \ac{PE} posterior samples.
$\xi(\veclams,\vecLam)$ is the expected fraction of events in the spacetime volume covered by the population prior and observing time that are detected. This is computed via Monte Carlo integration over an injection campaign of mock events \cite{2021RNAAS...5..220E,2018CQGra..35n5009T,2025PhRvD.112j2001E}.

The output of the Monte Carlo integration of a finite number of samples can deviate significantly from the true integral if the distribution of samples deviates significantly from the population distribution over which they are integrated \cite{2022arXiv220400461E,2023MNRAS.526.3495T}. To limit this error, we impose a threshold on the variance of the Monte Carlo log-likelihood estimator $\sigma^2_{\ln \mathcal{L}}$, which includes the errors from both the estimators of the individual event-level likelihoods and the selection function. Expressions for $\sigma^2_{\ln \mathcal{L}}$ corresponding to \cref{eq:hbi_ll} can be found in e.g. \cite{2025arXiv250907221H,2021RNAAS...5..220E}. During the inference, we set the likelihood to $0$ in regions of parameter space where $\sigma^2_{\ln \mathcal{L}} \geq 1$, following \cite{2023MNRAS.526.3495T}.

\subsection{Population models} \label{ssec:popmodels}

The results of the inference will naturally depend on the parametric form that one chooses for each $\di N_k/\di \vectheta$. In making this choice, one needs to strike a balance between imposing prior assumptions that are too strong (which could bias the results of the inference) and prior assumptions that are too weak (which could lead to overfitting of the data as well as computational challenges). Thus, we implement several variants of population models, which we describe below. A list of all of our population parameters across all population models, as well as the priors used during the inference, can be found in \cref{tab:params}.

\bmhead{Default model (\texttt{skewt})}

Inspired by the common phenomenological modeling of the primary mass distribution as a mixture model of power-law and Gaussian distributions \cite{2018ApJ...856..173T,2025arXiv250818083T}, we model the primary mass distribution as a Jones and Faddy skew-t distribution \cite{Jones2003}, which has the property of being able to asymptotically approach both tail-heavy power-law behavior as well as Gaussian behavior. It is given by
\begin{equation} \label{eq:jf_skew_t}
    \mathcal{S}(x; a, b) = C_{a,b}^{-1}
                    \left(1+\frac{x}{\left(a+b+x^2\right)^{1/2}}\right)^{a+1/2}
                    \left(1-\frac{x}{\left(a+b+x^2\right)^{1/2}}\right)^{b+1/2} 
\end{equation}
for real numbers $a>0$ and $b>0$, where $C_{a,b} = 2^{a+b-1}B(a,b)(a+b)^{1/2}$, and $B$ is the beta function. The general skew-t distribution has power-law behavior $\propto |x|^{2a+1}$ as $x \rightarrow \infty$ and $\propto |x|^{2b+1}$ as $x \rightarrow -\infty$; setting $a = b = \nu/2$ recovers the Student-t distribution with $\nu$ degrees of freedom. Subsequently taking the limit $\nu \rightarrow \infty$ recovers a standard Gaussian distribution.

We allow for a shift parameter $\mu_m$ and a scale parameter $\sigma_m$, evaluating \cref{eq:jf_skew_t} at
\begin{equation}
    x = \frac{m_1 - \mu_m}{\sigma_m},
\end{equation}
and truncate the distribution at a global \mmin as well as the maximum mass \mmax that we pin to $300\,\Msun$, following \cite{2025arXiv250818083T}. 
Finally, we re-parameterize and shift the distribution in \cref{eq:jf_skew_t} to improve the sampling efficiency. The final primary mass distribution we use is 
\begin{equation} \label{eq:m1_skewt_pdf}
    p_\mathcal{S}(m_1 | \log\alpha, \log\kappa, \mu_m, \sigma_m, \mmin, \mmax) \propto 
    \begin{cases}
        \frac{1}{\sigma_m}\mathcal{S}\left(\frac{m_1 - \mu_m}{\sigma_m} + \tilde{m}; \frac{\alpha\kappa}{1+\kappa}, \frac{\alpha}{1 + \kappa} \right) & \mmin \leq m_1 \leq \mmax \\
        0 & \text{otherwise}.
    \end{cases}
\end{equation}
where $a, b$ have been re-parameterized to a tail weight parameter $\alpha=a+b$ and skew parameter $\kappa=a/b$, and $\tilde{m}$ is the mode of \cref{eq:jf_skew_t}, given by
\begin{equation} \label{eq:skewt_mode}
    \tilde{m}(a,b) = \frac{(a - b) \sqrt{a + b}}{\sqrt{(2a + 1) (2b + 1)}}.
\end{equation}
With this parameterization, $\mu_m$ is the mode of the distribution, $\sigma_m$ measures the width of the peak, $\kappa$ characterizes the asymmetry of the power law tails, and $\alpha$ measures the tail weight, with large $\alpha$ approaching a Gaussian distribution.

We model the conditional mass ratio distribution as a truncated Gaussian with mean $\mu_q$ and width $\sigma_q$
\begin{equation} \label{eq:q_cond_pdf}
    p(q | m_1, \mu_q, \sigma_q, \mmin) = \mathcal{N}\left(q; \mu_q, \sigma_q, \frac{\mmin}{m_1}, 1\right),
\end{equation}
which allows for a peak in $p(q|m_1)$ away from $q=1$, motivated by possible formation mechanisms which favor unequal masses, including the \ac{SMT} channel \cite{2024A&A...689A.305O} and hierarchical mergers in \ac{AGN} disks or clusters \cite{2019PhRvL.123r1101Y,2020MNRAS.498.4088M,2017PhRvD..95l4046G}. Here and henceforth, $\mathcal{N}(x; \mu, \sigma, x_{\min}, x_{\max})$ will denote a Gaussian with mean $\mu$ and standard deviation $\sigma$, truncated on $x \in [x_{\min}, x_{\max}]$.

Finally, we use for all models a truncated Gaussian for the \chieff distribution, with mean $\mu_\chi$ and width $\sigma_\chi$
\begin{equation}
p(\chieff | \mu_\chi, \sigma_\chi) = \mathcal{N}(\chieff; \mu_\chi, \sigma_\chi, -1, 1).
\end{equation}


\bmhead{Power-laws and peaks (\texttt{NPLNP})}

In this model, a subpopulation's primary mass distribution is given by either a power-law distribution or a Gaussian. This allows us to compare our results more directly to phenomenological analyses in the literature. In this framework, the primary mass distribution is then a sum of $n_\text{PL}$ power-laws and $n_\mathcal{G}$ Gaussians, where $1 \leq n_\text{PL}, n_\mathcal{G} \leq 2$
\footnote{In preliminary analyses we have found that more than $2$ of either component is strongly disfavored, and thus for better convergence of our final result we limited the sampler to a maximum of $2$ of each component.}.

Our implementation of a power-law distribution with low-mass tapering is slightly different from the standard form used by the \ac{LVK}. Instead of a Planck tapering function, we use a power-law smoothing parameter $p$, making our probability distribution analytically normalizable:
\begin{equation} \label{eq:m1_PL_pdf}
    p_\text{PL}(m_1 | \alpha, p, \mmin^\PL, \mmin, \mmax^\PL) = 
    \begin{cases}
        \frac{1}{C} m_1^{-\alpha} \left(1 - \frac{\tilde{m}_{\min}}{m_1} \right)^p & \tilde{m}_{\min} \leq m_1 \leq \mmax^\PL \\
        0 & \text{otherwise}.
    \end{cases}
\end{equation}
Here, $-\alpha$ is the usual power-law index, and each subpopulation is allowed its own $\mmin^\PL$, and thus the distribution is smoothed to $0$ at
\begin{equation}
    \tilde{m}_{\min} = \max\left[ \mmin, \mmin^\PL \right].
\end{equation}
where $\mmin$ is the global \ac{BBH} minimum mass. 
Additionally, each subpopulation also has its own $\mmax^\PL \leq \mmax$. This effectively allows for a broken power-law as a sum of two power-laws with non-overlapping bounds. As in the default model, the global $\mmax$ is pinned at $300\,\Msun$.

The normalization of \cref{eq:m1_PL_pdf} is given by 
\begin{equation} \label{eq:m1_PL_norm}
    C = \left(\tilde{m}_{\min}\right)^{1-\alpha} I_{1-\tilde{m}_{\min}/\mmax^\PL}(p+1, \alpha-1)
\end{equation}
where $I_u(a, b)$ is the incomplete beta function. 
Note that the normalization here is valid only for $\alpha > 1$; computing the normalization for $\alpha \leq 1$ involves hypergeometric functions, and thus for computational convenience, we exclude $\alpha \leq 1$ from the prior. 

The Gaussian component is given by
\begin{equation} \label{eq:m1_gauss_pdf}
    p_\mathcal{G}(m_1 | \mu_q, \sigma_q, \mmin) = \mathcal{N}(m_1; \mu_m, \sigma_m, \mmin, \mmax),
\end{equation}
and therefore the total \ac{BBH} population distribution for $n_\PL$ power-law components and $n_\mathcal{G}$ Gaussian components is given by 
\begin{equation}
    \begin{split}
        \deriv{N}{\vectheta}(\veclams, \vecLam) = 
        T_\text{obs} (1+z)^{\gamma-1} \deriv{V_c}{z} 
        \bigg[ \sum_{k=1}^{n_\PL} &\mathcal{R}_0^k\, p_\PL(m_1 | \veclam^k_{m, \PL}, \mmin, \mmax)\,p(q | m_1, \veclam^k_q, \mmin)\,p(\chieff | \veclam^k_\chi) + \\
        \sum_{\ell=1}^{n_\mathcal{G}}  &\mathcal{R}_0^\ell\, p_\mathcal{G}(m_1 | \veclam^\ell_{m, \mathcal{G}}, \mmin, \mmax)\,p(q | m_1, \veclam^\ell_q, \mmin)\,p(\chieff | \veclam^\ell_\chi) \bigg].
    \end{split}
\end{equation}
where $\veclam^k_\theta$ are the subpopulation parameters corresponding to distributions $p_k(\theta)$ given above, and $\vecLam=[\mmin, \gamma]$ are global population parameters.

\bmhead{Identical mass distributions (\texttt{skewt\_ID})}

In general, it may be astrophysically more meaningful to model the distribution of \textit{all} merging black holes, rather than the more massive one of the two in the binary, which can be obscured by both sorting uncertainties when the mass ratio is close to unity \cite{2025PhRvL.134l1402G,2021PhRvL.126q1103B} as well as astrophysical uncertainties (e.g., the primary mass of the \ac{BBH} may not be the first born or the primary mass at zero age main sequence, due to mass ratio reversal \cite{2022ApJ...938...45B,2022MNRAS.517.2738M}). 
Thus, as an alternative, suppose both \ac{BBH} masses are drawn from the same underlying distribution $m_A, m_B \sim f(m)$ with \ac{CDF} $F(m)$, such that $m_1 = \max(m_A, m_B)$ and $m_2 = \min(m_A, m_B)$. Generically, we can write the joint probability distribution function as
\begin{equation} \label{eq:ID_pdf}
p(m_1, m_2) = 2 f(m_1) f(m_2) c(F(m_1), F(m_2))
\end{equation}
where $c$ is a copula describing the pairing between the two masses.
The factor of 2 accounts for the two possible orderings of $(m_A, m_B)$.

A simple choice for $c$ is the bivariate Gaussian copula density with correlation parameter $\rho \in (-1, 1)$:
\begin{equation}
c(u, v; \rho) = \frac{1}{\sqrt{1-\rho^2}} 
\exp \left(\frac{2 \rho z_1 z_2-\rho^2 (z_1^2+z_2^2)}{2(1-\rho^2)}\right),
\end{equation}
where $z_i = \Phi^{-1}(u_i)$, with $\Phi$ as the standard normal \ac{CDF} (and $\Phi^{-1}$ its inverse).
The marginal distributions for $m_1$ and $m_2$ can be shown to be 
\begin{align}
    p(m_1 | \rho) &= 2 f(m_1) \,\Phi\left(\sqrt{\frac{1-\rho}{1+\rho}}\,\Phi^{-1}\left(F(m_1)\right)\right) \\
    p(m_2 | \rho) &= 2 f(m_2) \,\Phi\left(-\sqrt{\frac{1-\rho}{1+\rho}}\,\Phi^{-1}\left(F(m_2)\right)\right).
\end{align}
We note that the marginals on the unordered masses, $(m_A, m_B)$ are $(f(m_A), f(m_B))$ by the usual properties of a copula, but the marginals are modified when we take the max/min. For $\rho = 0$, the above expressions become $2f(m_1) F(m_1)$ and $2 f(m_2) (1-F(m_2))$, which are the usual pdfs of the max/min of independent-identically-distributed random variables.
These expressions show how the correlation parameter $\rho$ shifts the marginal distributions: positive $\rho$ increases the probability of similar masses (both high or both low), while negative $\rho$ favors dissimilar masses. At the extremes, 
\begin{align}
p\left(m_2 \mid m_1, \rho=1 \right) &= \delta^D(m_2 - m_1) \\
p\left(m_2 \mid m_1, \rho=-1\right) &=  \delta^D\left(m_2-F^{-1}\left(1-F\left(m_1\right)\right)\right)
\end{align}
where $\delta^D$ is the Dirac-delta function. For $\rho=1$, the two masses are always equal, while at $\rho=-1$, $m_2$ is always at the quantile opposite of $m_1$ in the distribution $f(m)$.

Because copula models have the advantage of preserving the marginals of the two parameters, they have been used in past \ac{GW} population analyses to search for correlations such as $q-\chieff$ \cite{2022MNRAS.517.3928A,2023ApJ...958...13A,2025arXiv250818083T}, but not the pairing of \ac{BH} masses. 
References \cite{2020ApJ...891L..27F,2024ApJ...962...69F} analyzed the primary and secondary mass distributions of the \ac{BBH} population, but with a power-law pairing function rather than a copula. This is therefore a novel population model that probes the underlying \ac{BH} mass function of merging binaries and its binary pairings.

In the \texttt{skewt\_ID} model, we model each subpopulation $k$ as having its own correlation coefficient $\rho_k$ and mass distribution $f_k(m)=p_\mathcal{S}(m)$, which we model using the skew-t distribution given in \cref{eq:m1_skewt_pdf}.

\bmhead{Local redshift evolution (\texttt{skewt\_z})}

In this model, we relax the assumption of a global rate evolution and allow each subpopulation $k$ to have its own power-law rate index $\gamma_k$; it is otherwise identical to \texttt{skewt}.

\bmhead{Secondary mass maximum (\texttt{skewt\_m2max})}

Recently, \cite{2025arXiv250904151T} has found evidence for a gap in the secondary mass distribution, consistent with the theoretically predicted gap in the \ac{BH} mass function from \acp{PISN}. We verify these results by employing a variant of our default population model in which the conditional mass ratio distribution is truncated at a secondary mass maximum $m_{2, \max}$, which we infer as a global population parameter:
\begin{equation} \label{eq:q_cond_m2max_pdf}
    p(q | m_1, \mu_q, \sigma_q, \mmin, \mtmax) = \mathcal{N}\left(q; \mu_q, \sigma_q, \frac{\mmin}{m_1}, \min\left[1, \frac{\mtmax}{m_1}\right]\right).
\end{equation}
This targets the steep drop-off of black holes in the secondary mass distribution beyond $\sim45\,\Msun$ found by Tong et al. \cite{2025arXiv250904151T}. They find a gap in the secondary mass distribution, parameterizing the mass ratio distribution in a similar fashion to \cref{eq:q_cond_m2max_pdf}. 
Because the only event which populates the other side of the gap is GW231123, an analysis which allows for a secondary mass maximum and excludes GW231123 should find a preference for a maximum consistent with the lower edge of the gap.
Thus, we run \texttt{skewt\_m2max} both including and excluding GW231123 from the analysis. We show the results for this analysis in \cref{fig:m2max}.

\subsection{Reversible-jump MCMC} \label{sec:rjmcmc}

\ac{RJ MCMC}~\cite{1995Biome..82..711G} provides a powerful framework for exploring parameter spaces whose dimensionality is itself unknown. In addition to proposing updates to parameters within a given model, \ac{RJ MCMC} allows transitions between models of different dimensionality by introducing or removing parameters. The update of parameters for a fixed dimensionality happens as in a regular MCMC. We now describe how updates to the number of components are performed.

As before, let $\veclams_n=(\veclam_1,\veclam_2,...,\veclam_n)$ be the hyperparameters of the $n$ components of the model that are added or removed when adding or removing components, and $\vecLam$ be the other hyperparameters.
When proposing to add one component to a model with $n$ components, the acceptance ratio is $\alpha = \min[1, H_{n \to n+1}]$ with Hastings ratio \cite{2020PhRvD.101l3021L}
\begin{align} 
H_{n \to n+1} &= \frac{p(\{d\}|\veclams_{n+1},\vecLam|) \, \pi(\veclams_{n+1}) \,\pi(n+1) } 
{p(\{d\}|\veclams_{n},\vecLam|) \, \pi(\veclams_{n}) \, \pi(n) \, q(\veclam_{n+1})} \\
&= \frac{p(\{d\}|\veclams_{n+1},\vecLam|)} 
{p(\{d\}|\veclams_{n},\vecLam|)} \frac{\pi(\veclam_{n+1}) }{q(\veclam_{n+1})} \label{eq:acc}
\end{align}
where $\pi(\veclam)$ is the prior on $\veclam$, $\pi(n)$ are the prior odds on a model with $n$ components (which we take to be uniform across $n$), and $q(\veclam)$ is the distribution from which a new $\veclam$ is proposed. 

For a fixed proposal distribution $q(\veclam)$, a broader prior $\pi(\veclam)$ suppresses the second term in \cref{eq:acc}, implying that the newly proposed component must yield a larger increase in likelihood to be accepted. Conversely, proposals that are tightly concentrated in high-likelihood regions are penalized when they are much more localized than the prior. This behavior provides a direct implementation of Occam’s razor: models with additional components are only favored if they improve the fit to the data sufficiently to compensate for the increase in prior volume. In this work, we use the prior distribution itself as the proposal function, $q(\veclam)=\pi(\veclam)$.
In this case, the second term reduces to unity, but the Occam penalty is still present: as the prior volume increases, it becomes progressively harder to propose a new component that lands in a region of sufficiently high likelihood to be accepted. We find this choice to be a good compromise for our purpose. The acceptance ratio for removing a component is the inverse of \cref{eq:acc}, and the same considerations regarding the choice of prior and proposal apply.


\ac{RJ MCMC} enables efficient Bayesian model selection: the ratio of samples corresponding to models with $n_1$ and $n_2$ components directly estimates the Bayes factor between these models. 

\subsection{Implementation and sampling details} \label{ssec:implementation}

We analyze the 153 \acp{BBH} in \ac{GWTC-4} \cite{2025arXiv250818082T} with False Alarm Rate $< 1\,\mathrm{yr}^{-1}$, consistent with \cite{Collaboration2025}.
For events from O4a, we use \ac{PE} samples \cite{Collaboration2025} with the \texttt{NRSur7dq4} waveform where available, and \texttt{Mixed} samples otherwise. For earlier events, we use \ac{PE} samples \cite{Collaboration2023} with the \texttt{IMRPhenomXPHM} waveform.
For the \ac{RJ MCMC}, we employ the implementation in \texttt{eryn} ~\cite{2023MNRAS.526.4814K}. To improve computational efficiency, we have implemented our pipeline on GPUs; the analyses presented here require $\approx 10$ hours for a $60,000$ step run (including the burn-in) on a single GPU core.

The likelihood surface of a mixture model is highly degenerate, especially with respect to the rates of subpopulations: increasing the rate of one subpopulation naturally leads to a decrease in the rate of another subpopulation along level surfaces of the log-likelihood.
Therefore, we implement two custom Gibbs proposal steps to traverse this likelihood surface, and use the default Gibbs Gaussian proposal for the remaining parameters. 

The first custom move is an \emph{orthogonal proposal} that updates the local merger rate vector $\bm{\mathcal{R}} \equiv [\mathcal{R}_{0, 1}, \mathcal{R}_{0, 2}, \ldots, \mathcal{R}_{0, n}]$ of the $n$ subpopulations within the hyperplane orthogonal to $\mathbf{1}$, such that the total rate $\sum_k{\mathcal{R}_{0, k}}$ is conserved. Concretely, we draw $\mathbf{\Delta}\sim\mathcal{N}(\mathbf{0}, \sigma_\mathcal{R}^2\mathbf{1})$, where $\sigma_\mathcal{R}$ is a suitable proposal scale (which we take to be $\sigma_{\mathcal{R}}=1\,\mathrm{Gpc}^{-3}\,\mathrm{yr}^{-1}$), and perturb the walker by
\begin{equation}
\delta \bm{\mathcal{R}}
= \mathbf{\Delta} - \text{mean}[\mathbf{\Delta}],
\end{equation}
where $\text{mean}[\mathbf{z}]$ is the mean of $\mathbf{\Delta}$ over its $n$ components. 

Similarly, the second move is a \emph{radial proposal} parallel to the $\mathbf{1}$ direction, with
\begin{equation}
\delta \bm{\mathcal{R}} = \Delta \cdot\,\mathbf{1}, \qquad \Delta \sim \mathcal{N}(0,\sigma_\mathcal{R}^2),
\end{equation}
so that all rate components change by the same amount. This move changes the total rate without altering the relative rates between subpopulations. Alternating between these two Gibbs proposals therefore separates the redistribution of branching fraction among subpopulations from the overall rate normalization, improving mixing relative to a diagonal Gaussian proposal.

\subsection{Population-weighted posteriors and probabilities} \label{ssec:popweights}

In \cref{fig:subpop_prob}, we showed the probabilities of subpopulation membership of various events in the catalog; in \cref{fig:skewt_pwp,fig:skewt_z_pwp}, we showed contours of their population-weighted posteriors, i.e. the posterior probability of the source parameters $\vectheta_i$ of event $i$ given the data $\{d\}$. We derive expressions for these below.

Let $k=1,2,\ldots,n$ index the subpopulations of a given model (of fixed dimensionality), and $k_i$ denote the membership of event $i$ to subpopulation $k$.
From Bayes' theorem, the probability that event $i$ belongs to subpopulation $k$ conditioned on the data and population parameters $\veclams, \vecLam$ is given by
\begin{equation} \label{eq:post_ki}
    p(k_i | \{d\}) \propto \int \di \veclams \di \vecLam \,\mathcal{L}(d_i | k_i, \veclams, \vecLam) \, p(k_i | \veclams, \vecLam) \, p(\veclams, \vecLam | \{d\}_{\neq i}).
\end{equation}
$p(\veclams, \vecLam | \{d\}_{\neq i})$ is the ``leave-one-out'' posterior probability of the hyperparameters that condition on the data of all events except for event $i$, to avoid double counting. This is related to the posterior probability of the full analysis via
\begin{equation} \label{eq:post_leaveoneout}
    p(\veclams, \vecLam | \{d\}_{\neq i}) \propto  \frac{p(\veclams, \vecLam | \{d\})}{\mathcal{L}(d_i | \veclams, \vecLam)}
\end{equation}
taking advantage of the fact that the event-level likelihoods factorize.
In \cref{eq:post_ki}, $p(k_i | \veclams, \vecLam)$ is the probability of event $i$ belonging to subpopulation $k_i$ given the population; it has no dependence on $\{d\}_{\neq i}$ since it is conditionally independent of the other events' data given the population parameters. It is given by
\begin{equation} \label{eq:p_ki}
    p(k_i | \veclams, \vecLam) = \frac{N_{k_i}(\veclam_{k_i}, \vecLam)}{N(\veclams, \vecLam)}
\end{equation}
where $N_k(\veclam_k, \vecLam)$ is the total expected number of mergers of subpopulation $k$, obtained by integrating its differential merger rate over \vectheta.

The likelihood of $d_i$ given that it belongs to subpopulation $k_i$ and population parameters is 
\begin{equation} \label{eq:ll_di_given_ki}
    \mathcal{L}(d_i | k_i, \veclams, \vecLam) = \int \di \vectheta \,  \frac{\mathcal{L}(d_i | \vectheta)}{N_{k_i}(\veclam_{k_i}, \vecLam)} \frac{\mathrm d N_{k_i}}{\mathrm d\vectheta}(\veclam_{k_i},\vecLam).
\end{equation}
This can be approximated as a Monte Carlo sum over event $i$'s posterior samples $\vectheta^p_i$ 
\begin{equation} \label{eq:ll_di_given_ki_mc}
    \mathcal{L}(d_i | k_i, \veclams, \vecLam) \appropto \sum_p \frac{1}{\pi_{\rm PE}(\vectheta^p_i)} \frac{1}{N_{k_i}(\veclam_{k_i}, \vecLam)} \frac{\mathrm d N_{k_i}}{\mathrm d\vectheta}(\veclam_{k_i},\vecLam).
\end{equation}
\cref{eq:post_ki} can finally be computed by multiplying together \cref{eq:post_leaveoneout,eq:p_ki,eq:ll_di_given_ki_mc} and expressing the integral as a sum over population posterior samples $\{\veclam^m\}, \vecLam^m$ 
\begin{equation}
    p(k_i | \{d\}) \appropto  \sum_p \frac{1}{\pi_{\rm PE}(\vectheta^p_i)} \sum_m \frac{\mathrm d N_{k_i}}{\mathrm d\vectheta}( \veclam_{k_i}^m,\vecLam^m)\Big|_{\vectheta^p_i}  \Big[ N(\{\veclam^m\}, \vecLam^m)\,\mathcal{L}(d_i | \{\veclam^m\}, \vecLam^m)\Big]^{-1}
\end{equation}
which is what we used to compute the probabilities in \cref{fig:subpop_prob}. The bracketed quantities can be pre-computed for each population posterior sample $m$; $\mathcal{L}(d_i | \{\veclam^m\}, \vecLam^m)$ can also be computed via Monte Carlo integration, marginalizing \cref{eq:ll_di_given_ki} over the different subpopulations
\begin{equation}
    \mathcal{L}(d_i | \veclams, \vecLam) \appropto \sum_p \frac{1}{\pi_{\rm PE}(\vectheta^p_i)} \frac{1}{N(\veclams, \vecLam)} \sum_{k}  \frac{\mathrm d N_{k}}{\mathrm d\vectheta}(\veclam_k,\vecLam).
\end{equation}

Next, the population-weighted posterior
\begin{equation}
    p(\vectheta_i | \{d\}) \propto \mathcal{L}(d_i | \vectheta_i) p(\vectheta_i | \{d\}_{\neq i})
\end{equation}
can be obtained in a similar manner, following ref. \cite{2020PhRvD.102h3026G}:
\begin{equation}
    p(\vectheta_i | \{d\}) \appropto  \frac{p(\vectheta_i | d_i)}{\pi_{\rm PE}(\vectheta_i)} \sum_k \sum_m \frac{\mathrm d N_{k}}{\mathrm d\vectheta}( \veclam_k^m, \vecLam^m) \Big|_{\vectheta_i} \Big[ N(\{\veclam^m\}, \vecLam^m)\,\mathcal{L}(d_i | \{\veclam^m\}, \vecLam^m)\Big]^{-1}.
\end{equation}
In practice, we assign each posterior sample $\vectheta_i^p$ a weight
\begin{equation}
w_{i}^p \propto
\frac{1}{\pi_{\rm PE}(\vectheta_i^p)} \sum_k \sum_m \frac{\mathrm d N_{k}}{\mathrm d\vectheta}( \veclam_k,\vecLam)\Big|_{\vectheta^p_i} \Big[ N(\{\veclam^m\}, \vecLam^m)\,\mathcal{L}(d_i | \{\veclam^m\}, \vecLam^m)\Big]^{-1}
\end{equation}
from which we can use a kernel density estimator to create the contours in \cref{fig:skewt_pwp,fig:skewt_z_pwp}.

\subsection{Subpopulation uniqueness} \label{ssec:unique}

Because our goal is to search for distinct subpopulations in parameter space, one must penalize the sampler against fitting the data with multiple close-together or identical components when it can be fit with a single larger component. Furthermore, one also needs to be able to identify the different components found by the sampler in post-processing across all of the posterior samples. 
We do so by mapping for each subpopulation $k$ the distribution $p_k(\theta | \veclam_k, \vecLam)$ of each parameter $\theta$ to a whitened feature space $\mathbf{x}_k$ roughly corresponding to the first two moments of the distribution. This mapping is used to ensure that found components are distinct during inference, as well as to sort components in post-processing into identified groups, or subpopulations.

We now describe this mapping $p \to \mathbf{x}$ as follows.
For some \ac{BBH} parameter $\theta \in [m_1, q, \chieff]$, each component $k$ has its own probability distribution $p_k(\theta | \veclam_k, \vecLam)$. We map this distribution into a feature space by computing a characteristic center $c_k^\theta(\veclam_k, \vecLam)$ and width $w_k^\theta(\veclam_k, \vecLam)$; we describe the formulae for these in \cref{ssec:moments}. Because these quantities are in general dimensionful with different characteristic scales, we used the whitened vector
\begin{equation} \label{eq:feature_space}
    \mathbf{x}_k(\veclam_k, \vecLam) = \left[ \frac{c^{m_1}_k}{\bar{m}}, \frac{w^{m_1}_k}{\bar{m}}, \frac{c^{q}_k}{\bar{q}}, \frac{w^{q}_k}{\bar{q}}, \frac{c^{\chieff}_k}{\bar{\chi}_\mathrm{eff}}, \frac{w^{\chieff}_k}{\bar{\chi}_\mathrm{eff}}\right],
\end{equation}
rescaling each $c^\theta_k, w^\theta_k$ by a characteristic whitening scale $\bar{\theta}$.
Note that for $p_k(q)$, we are using the conditional mass ratio distribution in \cref{eq:q_cond_pdf}, except we take the truncation between $0$ and $1$ to remove the $m_1$ dependence for computational expediency. For the copula model \texttt{skewt\_ID}, we compute the features of the total \ac{BBH} mass distribution $f_k(m)$ rather than $p_k(m_1)$ and $p_k(q)$. For \texttt{skewt\_z}, we add in the independent redshift evolution of each component by adding the redshift power-law index $\gamma$ itself to the feature vector, similarly rescaled by a characteristic $\bar{\gamma}$.

In post-processing, a natural whitening scale $\bar{\theta}$ can be defined from both the standard deviation of the features used for clustering and the intrinsic standard deviation of each component. Specifically, we define the whitening scales as
\begin{align} \label{eq:auto_whiten}
    \bar{\theta} &= \sqrt{\mathrm{Var}[c^\theta] + \mathrm{Var}[w^\theta] + \mathrm{median}[w^\theta]^2} \qquad\quad \theta \in [m_1, q, \chieff] \\ 
    \bar{\gamma} &= \sqrt{\mathrm{Var}[\gamma]},
\end{align}
where sample median and variances are taken across the $c^\theta_k(\veclam_k^m, \vecLam^m)$, $w^\theta_k(\veclam_k^m, \vecLam^m)$, $\gamma_k^m$ computed from the population draw corresponding to each posterior sample $\veclam^m_k, \vecLam^m$. This choice ensures that components are considered distinct only when their separation is larger than the corresponding statistical uncertainty and when the components themselves are sufficiently well separated to be distinguishable.

With the whitening scale computed with \cref{eq:auto_whiten}, we can now map each posterior sample output by the \ac{RJ MCMC} to a feature vector as in \cref{eq:feature_space}, $p_k(\vectheta | \veclam_k^m, \vecLam^m) \to \mathbf{x}^k_m$.
We then sort the components by performing $k$-means clustering on these feature vectors, using the implementation in \texttt{scikit-learn} \cite{scikit-learn}.
$k$-means clustering is a fast and simple clustering algorithm which finds clusters by initializing $n_c$ cluster centroids and minimizing the distance between data points and their cluster centroids; because of this, whitening of the feature space is especially necessary. The number of centroids $n_c$ must be provided by the user.

\ac{RJ MCMC} returns posterior samples for all models being compared simultaneously, with the number of samples in each model proportional to the model evidences and therefore the Bayes factor. To identify corresponding subpopulations across different posterior samples, we apply the clustering algorithm to all inferred components within a given model family (e.g. power-laws, Gaussians or skew-ts), combining samples with different numbers of components $n$. In this way, components that appear in different posterior samples can be grouped into common clusters representing the same underlying subpopulation.
Because the $k$-means algorithm requires the number of clusters, $n_c$, to be specified in advance, we set $n_c$ equal to the largest number of components among all models that are not strongly disfavoured relative to the preferred model. Specifically, we consider all models with a Bayes factor $\mathcal{B} \geq 0.1$ with respect to the model with the highest evidence.
The output of the clustering algorithm is what we use to identify subpopulations. 

As discussed above, we also use \cref{eq:feature_space} to enforce the uniqueness of components during inference. During the inference, we do not have access to the posterior distribution of $c^\theta_k, w^\theta_k$, needed to compute the whitening scale defined in Eq.~\ref{eq:auto_whiten}, but, we still have a handle on the relevant scales of our different quantities. We use $\bar{m}=8\,\Msun$, $\bar{q}=0.3$, $\bar{\chi}_\mathrm{eff}=0.1$, and $\bar{\gamma}=1.0$ to compute the $\{\mathbf{x}_k\}$ corresponding to the position of the walker. These values were chosen to match those of \cref{eq:auto_whiten} in earlier test runs.
We then enforce the separation of subpopulations by setting the likelihood to $0$ when 
\begin{equation}
    \max\left[\mathrm{abs}[\mathbf{x}_{k^\prime} - \mathbf{x}_{k}]\right] < 1,
\end{equation}
where the maximum is taken over the different vector elements. Thus, the likelihood is set to $0$ when any two subpopulations are separated by less than $1$ along all dimensions in the feature space of $\mathbf{x}$.

Although the results are not particularly sensitive to the exact choice of $\bar{\theta}$, ill-chosen $\bar{\theta}$ can cause in confusion in interpreting the results. For example, using a sufficiently small $\bar{q}$ will cause poor sorting of subpopulations, since it effectively upweights the $w^q, c^q$ dimensions of the feature space when the $p(q)$ are not particularly well constrained. 
Furthermore, $\bar{\theta}$ (and especially $\bar{m}$) that is too small allows the sampler to explore 3-component models consisting of two identical components, which should in principle be counted as a 2-component model.
Although such cases are disfavored by the Occam's razor penalty, their posterior support is small but non-negligible for $\bar{m}=5\,\Msun$.

For population models with multiple model types, i.e., the \texttt{NPLNP} population model, there can still be confusion \textit{between} model types. For example, in the most preferred ``1 power-law + 2 Gaussians'' model of \texttt{NPLNP}, both the $10\,\Msun$ feature and the $30\,\Msun$ feature are Gaussians in $m_1$. However, both features can also be modeled with steep power-law distributions, and hence the ``2 power-laws + 1 Gaussian'' posterior is bimodal, with one power-law as the continuum, and the other power-law split between the $10\,\Msun$ and $30\,\Msun$ feature with about equal preference. Thus, upon comparing this result with \cref{tab:bfs}, this means that both subpopulations prefer to be represented with a Gaussian with $\BF \approx 3$.

\subsection{Formulae for computing $c$ and $w$} \label{ssec:moments}

\noindent \textbf{Truncated Gaussian}, $\mathcal{N}(x; \mu, \sigma, x_{\min}, x_{\max})$. 
\begin{align}
    c &= \mu \\
    w &=\dfrac{F^{-1}(0.84; \mu, \sigma, x_{\min}, x_{\max}) - F^{-1}(0.16; \mu, \sigma, x_{\min}, x_{\max})}{2}
\end{align}
where $F^{-1}$ is the inverse \ac{CDF} of the truncated Gaussian. In terms of the standard normal \ac{CDF} $\Phi(x)$, this is
\begin{equation}
\begin{split}
F^{-1}(q) &= \mu + \sigma\, \Phi^{-1}\!\left[q_{\rm lo} + q\,(q_{\rm hi} - q_{\rm lo})\right] \\
    q_{\rm lo} &= \Phi\!\left((x_{\min} - \mu)/\sigma\right) \\
    q_{\rm hi} &= \Phi\!\left((x_{\max} - \mu)/\sigma\right)
\end{split}.
\end{equation}
The truncated Gaussian is used to model the conditional mass ratio distribution (where for simplicity $c,w$ are evaluated at $x_{\min}=0$, $x_{\max}=1$) as well as the \chieff distribution. It is also used to model $m_1$ in \texttt{NPLNP}.

\vspace{2ex}
\noindent \textbf{Power-law with low-mass smoothing}, $p_\text{PL}(m_1 | \alpha, p, \mmin^\PL, \mmin, \mmax^\PL)$. See \cref{eq:m1_PL_pdf}.
\begin{align}
    c &= x_{\min}(1 + p/\alpha) \\
    w &= \frac{1}{2}\left[F_{\rm PL}^{-1}(0.5) - c\right]
\end{align}
where $F_{\rm PL}^{-1}$ is the inverse \ac{CDF} for a truncated power-law distribution (ignoring the low-mass smoothing for simplicity)
\begin{equation}
    F_{\rm PL}^{-1}(q) = \left[(x_{\max}^{1-\alpha} - x_{\min}^{1-\alpha})\,q + x_{\min}^{1-\alpha}\right]^{1/(1-\alpha)}.
\end{equation}
This is also used for $m_1$ in \texttt{NPLNP}.

\vspace{2ex}
\noindent \textbf{Jones and Faddy skew-t distribution}, $p_\mathcal{S}(m_1 | \log\alpha, \log\kappa, \mu_m, \sigma_m, \mmin, \mmax)$. See \cref{eq:m1_skewt_pdf}.
\begin{align}
    c &= \mu \\
    w &= \sigma_m \left[\frac{(a+b)(a+b+1)^4}{(4ab+2(a+b)+1)^3}\right]^{1/2} \left[\frac{a+1/2}{(2a+1)^2} + \frac{b+1/2}{(2b+1)^2}\right]^{-1/2}
\end{align}
This is used for $m_1$ in all models except \texttt{NPLNP}.

\backmatter

\bmhead{Data Availability}

All software used for this work is made available on Github \footnote{\url{https://github.com/aqcheng/RJpop}}. The data underlying this article can be downloaded at ref. \cite{cheng2026}.

\bmhead{Acknowledgements}
We are thankful to T. Bruel, H. Quelquejay-Leclere, Daniel Holz, and Colm Talbot for their constructive comments and to Hui Tong for completing the internal \ac{LVK} review. 
A.Q.C. is currently supported by the Lowell Wood Endowed Fellowship of the Fannie and John Hertz foundation.
A.T. is supported by MUR Young Researchers Grant No. SOE2024-0000125, ERC Starting Grant No.~945155--GWmining, Cariplo Foundation Grant No.~2021-0555, MUR PRIN Grant No.~2022-Z9X4XS, Italian-French University (UIF/UFI) Grant No.~2025-C3-386, MUR Grant ``Progetto Dipartimenti di Eccellenza 2023-2027'' (BiCoQ), and the ICSC National Research Centre funded by NextGenerationEU. 
This material is based upon work supported by NSF's LIGO Laboratory which is a major facility fully funded by the National Science Foundation.
The computational work for this manuscript was carried out on the compute clusters Saraswati and Hypatia at the Max Planck Institute for Gravitational Physics in Potsdam.

\bmhead{Competing interests}
The authors declare that they have no competing financial interests.

\bibliography{bib}

\newpage
\begin{table}
\centering
\small
\begin{tabular}{p{3cm} c l c}

\toprule
 & Parameter & Description & Prior \\
\midrule

\multirow[c]{4}{=}{\centering All}
 & $m_{\min}$ & Global minimum \ac{BH} mass & $\mathcal{U}[3, 7]$ \\
 & $\mathcal{R}_0$ & Local merger rate of each subpopulation & $\mathcal{U}[0.05, 100]$ \\
 & $\mu_\chi$ & Mean of truncated Gaussian $p(\chieff)$ & $\mathcal{U}[-1, 1]$\\
 & $\sigma_\chi$ & Width of truncated Gaussian $p(\chieff)$ & $\mathcal{U}[0.03, 1]$ \\
\midrule

\multirow[c]{4}{=}{\centering \texttt{skewt}}
 & $\mu_m$ & Peak location of $p_{\mathcal{S}}(m_1)$ & $\mathcal{U}[6, 60]$ \\
 & $\sigma_m$ & Scale parameter of $p_{\mathcal{S}}(m_1)$ & $\mathcal{U}[1, 15]$ \\
 & $\log\alpha$ & Tail weight parameter of $p_{\mathcal{S}}(m_1)$ & $\mathcal{U}[-1, 2]$\\
 & $\log\kappa$ & Skewness parameter of $p_{\mathcal{S}}(m_1)$ & $\mathcal{U}[-2, 2]$\\
\midrule

\multirow[c]{6}{=}{\centering \texttt{NPLNP}}
 & $\alpha$ & Power-law index of $p_\PL(m_1)$ & $\mathcal{U}[1.1, 12]$\\
 & $p$ & Low-mass smoothing parameter of $p_\PL(m_1)$ & $\mathcal{U}[0.5, 5]$\\
 & $m_{\min}^{\mathrm{PL}}$ & $p_\PL(m_1)$ minimum & $\mathcal{U}[3, 50]$\\
 & $m_{\max}^{\mathrm{PL}}$ & $p_\PL(m_1)$ maximum & $\mathcal{U}[50, 300]$\\
 & $\mu_m$ & Mean of Gaussian $p_{\mathcal{G}}(m_1)$ & $\mathcal{U}[-1, 1]$ \\
 & $\sigma_m$ & Width of Gaussian $p_{\mathcal{G}}(m_1)$ & $\mathcal{U}[1, 10]$ \\
\midrule

\multirow[c]{2}{=}{\centering All except \texttt{skewt\_ID}}
 & $\mu_q$ & Mean of truncated Gaussian $p(q)$ & $\mathcal{U}[0.1, 1]$ \\
 & $\sigma_q$ & Width of truncated Gaussian $p(q)$ & $\mathcal{U}[0.1, 1]$ \\
 \midrule 
 
\multirow[c]{1}{=}{\centering \texttt{skewt\_ID}}
& $\rho$ & $m_1-m_2$ Gaussian copula correlation & $\mathcal{U}[-1, 1]$ \\
\midrule

\multirow[c]{1}{=}{\centering \texttt{skewt\_m2max}} & $m_{2,\max}$ & Global secondary mass maximum & $\mathcal{U}[40, 200]$ \\
\midrule

\multirow[c]{1}{=}{\centering All except \texttt{skewt\_z}}
 & $\gamma$ & Global redshift power law index & $\mathcal{U}[-6, 6]$ \\
\midrule 
 
\multirow[c]{1}{=}{\centering \texttt{skewt\_z}}
& $\gamma$ & Subpopulation redshift power law index & $\mathcal{U}[-6, 10]$ \\
\bottomrule
\end{tabular}
\caption{Population parameters of all of our population models, as well as their prior ranges in the inference. $\mathcal{U}$ denotes a uniform distribution; masses are quoted in $\Msun$ and rates in $\rm{Gpc}^{-3}\,\mathrm{yr}^{-1}$. \label{tab:params}}
\end{table}


\begin{table}
\centering
\small
\begin{tabular}{cccccc||cc}
\toprule
 Model & \texttt{skewt} & \texttt{skewt\_ID} & \texttt{skewt\_z} & \texttt{skewt\_m2max} & \texttt{skewt\_m2max}\footnotemark[3] & Model & \texttt{NPLNP} \\
\midrule
 $n=2$ & 0.77 & 0.37 & 1.6 & 0.62 & 1.4  & 1 PL, 1 gauss  & 0.08 \\
 $n=3$ & 1 & 1 & 1 & 1 & 1 &  1 PL, 2 gauss  & 1 \\
 &- &- &- &- &- & 2 PL, 1 gauss  & 0.62 \\
 $n=4$ & 0.01 & 0.01 & 0.02 & 0.01 & 0.01 & 2 PL, 2 gauss  & 0.02 \\
\bottomrule
\end{tabular}
\caption{\label{tab:bfs} Bayes factors between models of varying complexity within each of our population models, relative to the 1 PL + 2 Gaussian model of \texttt{NPLNP} (right two columns) and the 3-component model of all other population models.}
\end{table}

\footnotetext{Without GW231123.}

\newpage

\section{Extended Data Figures}

This section presents supplementary figures referenced throughout the main text.

\begin{extdatafigure}
\centering
\includegraphics[width=\textwidth]{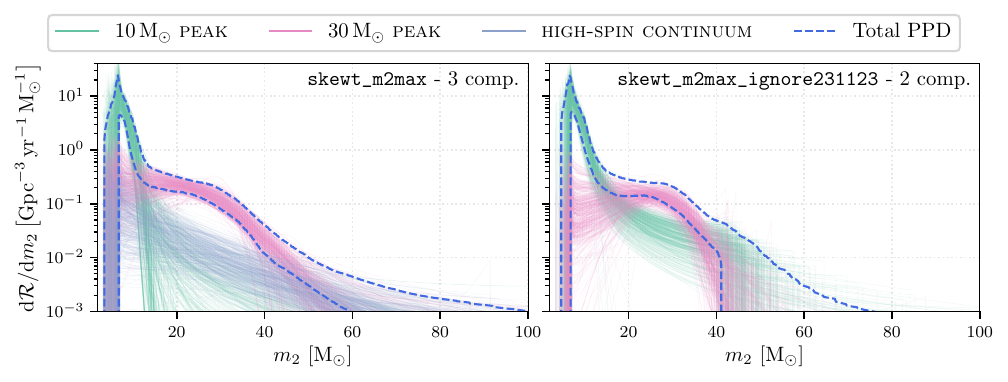}
\caption{Posterior draws from the marginal differential merger rate as a function of $m_2$ at $z=0.2$ for each subpopulation, inferred with (left) and without (right) GW231123. The $90\%$ credibility interval of the total rate is overplotted with the dashed blue lines. \label{fig:m2max}}
\end{extdatafigure}

In \cref{fig:m2max}, we show the differential merger rate in $m_2$ inferred from the \texttt{skewt\_m2max} population model allowing for a secondary mass maximum, including (left) and excluding (right) GW231123. As seen in the right panel, we do indeed find evidence for a steep drop-off around $50\,\Msun$ in the secondary-mass distribution when GW231123 is excluded, consistent with the \ac{PISN} gap. Nonetheless, the non-existence of a gap or maximum is not completely ruled out, with $m_{2,\max}=48^{+114}_{-7}~M_{\odot}$. 
Because GW231123 almost certainly belongs to the high-spin continuum subpopulation, excluding it from the inference is sufficient for the data to mildly prefer the simpler 2-component model over a 3-component model ($\BF=2.7$, compared to $\BF=0.6$ when GW231123 is included). This shows that individual events can significantly impact smaller subpopulations, such as the high-spin continuum, which constitutes only $\approx2\%$ of the underlying population.

Next, \cref{fig:all_ppds_other_models} shows the posterior marginals of $m_1$, $q$, and \chieff for the \texttt{NPLNP} and \texttt{skewt\_ID} population models. Together with \cref{fig:skewt_PPDs}, these results demonstrate the robustness of our three identified subpopulations across the different population models.
\Cref{fig:ppd_evolution} shows the inferred primary mass and $\chieff$ dsitributions at $z=0$, $0.5$, and $1$ under the \texttt{skewt\_z} model. 
Because the $10\,\Msun$ peak and continuum evolves faster than the $30\,\Msun$ peak, the latter becomes subdominant to the former at higher redshifts.
Furthermore, we find no evidence for the broadening of the \chieff distribution with redshift, as we discussed in \cref{sec:correlations}.
Finally, \cref{fig:q_chieff_contours} shows joint $q$--$\chieff$ contours of each subpopulation in the \texttt{skewt} population model, where we also find no significant $q$--$\chieff$ correlation.

\begin{extdatafigure}
\centering
\begin{subfigure}[t]{0.9\textwidth}
    \centering
    \includegraphics[width=\textwidth]{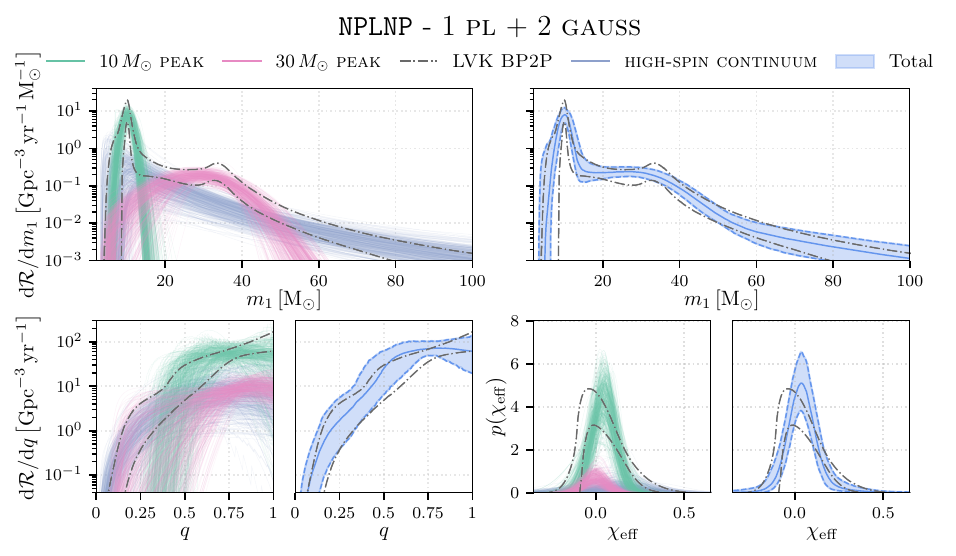}
    \caption{}
\end{subfigure}
\begin{subfigure}[t]{0.9\textwidth}
    \centering
    \includegraphics[width=\textwidth]{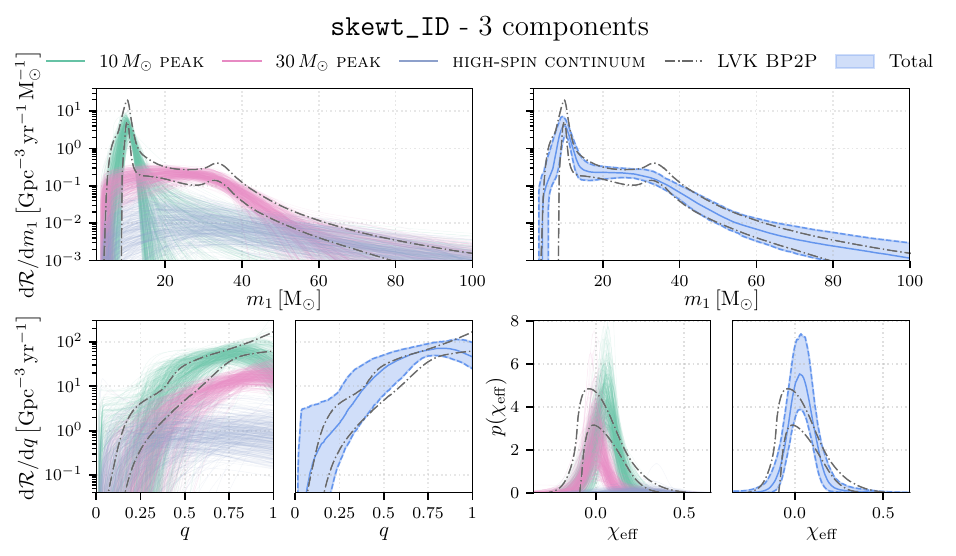}
    \caption{}
\end{subfigure}
\caption{\label{fig:all_ppds_other_models} The same as \cref{fig:skewt_PPDs} but for the \texttt{NPLNP} (a) and \texttt{skewt\_ID} (b) population models.}
\end{extdatafigure}

\begin{extdatafigure}
\centering
\begin{subfigure}[t]{0.493\textwidth}
    \centering
    \includegraphics[width=\textwidth]{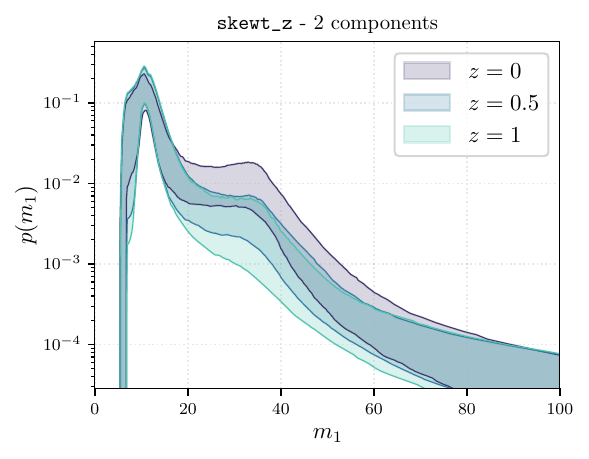}
    \caption{\label{fig:m1_evolution}}
\end{subfigure}
\hfill
\begin{subfigure}[t]{0.473\textwidth}
    \centering
    \includegraphics[width=\textwidth]{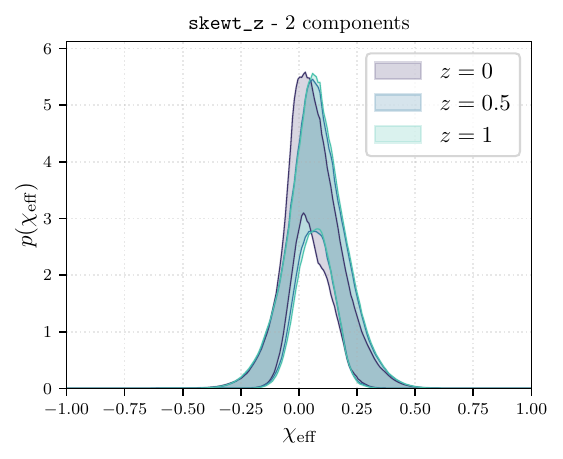}
    \caption{\label{fig:chieff_evolution}}
\end{subfigure}
\caption{\label{fig:ppd_evolution} 
$90\%$ credibility intervals on the marginal distributions of (a) $m_1$ and (b) \chieff at $z=0$, $z=0.5$, and $z=1$.}
\end{extdatafigure}

\begin{extdatafigure}
\centering
\includegraphics[width=0.7\textwidth]{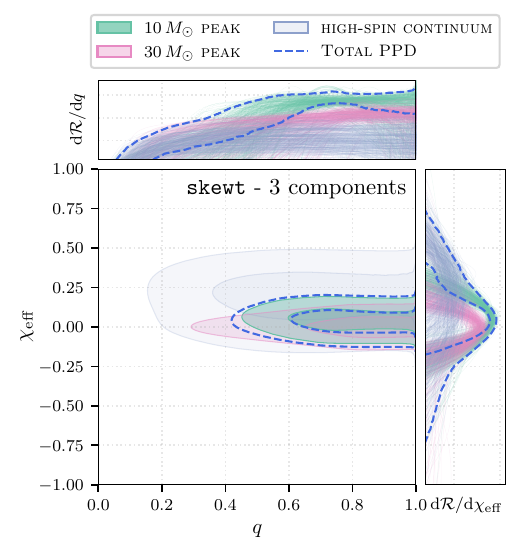}
\caption{\label{fig:q_chieff_contours} 
$50\%$ and $90\%$ contours of the median posterior $\di \mathcal{R}^k / \di q \, \di \chieff$ for each subpopulation $k$, evaluated at $z=0.2$ in the \texttt{skewt} model; opacity of the shading corresponds to the relative rates of the subpopulations. Corresponding posterior sample draws of the marginalized 1D distributions (in log-scale) are shown in the side panels.}
\end{extdatafigure}

\end{document}